	\documentclass[10pt, conference, letterpaper]{IEEEtran}

	\ifCLASSINFOpdf
	
	\else
	
	\fi
	
	\usepackage[linesnumbered,ruled]{algorithm2e}
	\usepackage{algpseudocode}
	\usepackage{bm}
	\usepackage{bbm}
	\usepackage{amssymb,amsthm,textcomp}
	
	\usepackage[fleqn]{amsmath}
	
	\usepackage{url}
	\usepackage[final]{pdfpages}
	\usepackage{listings}
	\usepackage{textcomp}
	\usepackage{color} 
	\usepackage{comment}
	
	\usepackage{amsthm}
	\usepackage{graphicx} 
	\usepackage{mathtools}
	
	\usepackage{MnSymbol} 
	\usepackage{multirow}

	\usepackage[section]{placeins}
	\usepackage[labelformat=simple]{subcaption}
	
	\usepackage{lipsum}
	\usepackage{paralist}
	\usepackage{wrapfig}
	
	\usepackage[font=small]{caption}
	\usepackage{cite}
	
	\makeatletter
	\newcommand\footnoteref[1]{\protected@xdef\@thefnmark{\ref{#1}}\@footnotemark}
	\makeatother

	\theoremstyle{remark}
	\newtheorem{definition}{Definition}[]

	\newtheorem{proposition}{Proposition}[]
	
	\newtheorem{remark}{Remark}[]
	\newtheorem{observation}{Observation}[]
	
	\usepackage [english]{babel}
	\usepackage [autostyle, english = american]{csquotes}
	\MakeOuterQuote{"}
	
	\addto\captionsenglish{}
	
	\definecolor{mygreen}{RGB}{28,172,0} 
	\definecolor{mylilas}{RGB}{170,55,241}
	
	\graphicspath{{./Figures/}}
	
\begin{document}
		
		\title{Smart Information Spreading for \\ Opinion Maximization in Social Networks}
		\author{\IEEEauthorblockN{Anuj Nayak, Seyyedali Hosseinalipour and Huaiyu Dai}
			\IEEEauthorblockA{Department of Electrical and Computer Engineering\\
				North Carolina State University\\
				Email: \{aknayak,shossei3,hdai\}@ncsu.edu}}
		
		\maketitle
		
		\IEEEpeerreviewmaketitle
		
		\begin{abstract}
			\textbf{
				The goal of opinion maximization is to maximize the positive view towards a product, an ideology or any entity among the individuals in social networks. So far, opinion maximization is mainly studied as finding a set of influential nodes for fast content dissemination in a social network. In this paper, we propose a novel approach to solve the problem, where opinion maximization is achieved through efficient information spreading. In our model, multiple sources inject information continuously into the network, while the regular nodes with heterogeneous social learning abilities spread the information to their acquaintances through gossip mechanism. One of the sources employs smart information spreading and the rest spread information randomly. We model the social interactions and evolution of opinions as a dynamic Bayesian network (DBN), using which the opinion maximization is formulated as a sequential decision problem. Since the problem is intractable, we develop multiple variants of centralized and decentralized learning algorithms to obtain approximate solutions. Through simulations in synthetic and real-world networks, we demonstrate two key results: 1) the proposed methods perform better than random spreading by a large margin, and 2) even though the smart source (that spreads the desired content) is unfavorably located in the network, it can outperform the contending random sources located at favorable positions.
			}
		\end{abstract}
		
		\section{Introduction}
		\noindent \textit{Opinion maximization} aims to maximize the affinity of individuals in a social network towards a specific product, political party or an ideology. It can manifest itself in various scenarios such as political campaigns \cite{lee2014social}, online marketing in social networks \cite{ellison2007benefits}, and advertisement dissemination in emerging networks such as VANETs \cite{hosseinalipour2017real, nayak2017dynamic}. Presently, social media platforms like Facebook, Twitter, etc., have been extensively used by the campaigners to form opinions through advertising. However, due to increasing advertising clutter, the campaigners witness advertising blindness from the users \cite{adblindness}. Consequently, word-of-mouth marketing seems to be one of the promising means of advertisement dissemination. According to the marketing research firm Nielsen, 92 percent of consumers around the world claim that they trust recommendations from friends and family members, above all other forms of advertising \cite{nielsen2012consumer}. Therefore, the aforementioned reasons motivate the need of campaigning methods that actively engage users in social networks, such as peer-to-peer advertising. In this context, gossip-based information exchange is a popular method to model peer-to-peer communications among the entities in large-scale distributed systems \cite{shah2009gossip}. In \cite{fernandess2008spreading}, the idea of social gossip is proposed for spreading recommendations in social networks. There are multiple works on peer-to-peer recommendation systems based on gossip protocols such as PREGO \cite{mordacchini2010p2p} and P2Prec \cite{draidi2011p2prec}.
		\subsection{Related Work}
        In \cite{gionis2013opinion}, opinion maximization in social networks is studied for the first time, where the objective is to find a subset of target individuals (seed nodes), whose positive opinion about a desired content maximizes the overall affinity towards it. Some heuristic algorithms, namely, freeDegree, RWR, etc., are proposed whose performances are evaluated in large-scale bibliographical datasets. Their approach is similar to the extensively studied \textit{influence maximization} problem \cite{kempe2003maximizing,borgs2014maximizing, tang2014influence,gomezinfluence,chen2009efficient,chen2010scalable,  chen2010linthresh, jung2012irie, kim2013scalable}, whose objective is to find seed nodes in a social network to be convinced to adopt a new product such that the number of individuals adopting the product in the long run (influence spread) by word-of-mouth spreading is maximized. In \cite{kempe2003maximizing}, a greedy hill-climbing algorithm to find the seed nodes is proposed, and is proved to achieve 63\% of the optimal influence spread. The techniques proposed in \cite{borgs2014maximizing, tang2014influence, gomezinfluence} are multiple low complexity versions of the greedy algorithm, with the influence spread close to the greedy algorithm. However, these are not scalable to large-scale networks, since their computational complexities are at least $O(E)$ ($E$ is the number of social links). To address the scalability issue of these algorithms, novel heuristic algorithms are proposed in \cite{chen2009efficient, chen2010scalable,  chen2010linthresh, jung2012irie, kim2013scalable}, and it is demonstrated through simulations that they achieve influence spread close to the greedy algorithm. A slightly different problem called topic-aware influence maximization is considered in \cite{chen2015online}, where the influence between any two individuals depends on topic-specific weights of social links. The objective is to find a subset of seed nodes that achieves the maximum influence spread for a given topic distribution. Multiple low complexity algorithms are proposed to compute approximate marginal influence spread due to each seed node, following which the exact influence spread is computed for only those nodes with larger marginal influence spread. In all the aforementioned works, the opinion or influence maximization is regarded as a problem of finding a subset of seed nodes. In \cite{sharara2011utilizing}, a different influence maximization problem is considered, where at every time step, a network operator needs to control the number of newly influenced users that utilize the network bandwidth (by downloading a file from a server, say). At any given time, it is achieved by switching on a fixed number of social links along which influence propagates. An algorithm is proposed to achieve maximum influence spread by strategically switching on the social links at every time step. However, the focus of our work is different from the aforementioned papers: we study the problem of opinion maximization from the angle of smart information spreading, which is described in the rest of the paper.
		
		\subsection{Our Contributions}	
		In this paper, we propose a new approach of efficient information spreading to address opinion maximization in social networks. First, we model the social interactions and opinion dynamics in the social network as a dynamic Bayesian network, using which opinion maximization is formulated as a sequential decision problem. Owing to its intractability, we provide a series of approximations to develop iterative centralized algorithms. Considering the scalability issue of the centralized algorithms, we further propose low-complexity online decentralized algorithms. Through simulations, we demonstrate the effectiveness of our algorithms on both the preferential attachment (PA) graphs and the Facebook ego-network \cite{snapnets},\cite{leskovec2012learning}, which is a snapshot of a real social network. To the best of our knowledge it is for the first time that the opinion maximization is formulated and studied as an information spreading problem.
		
		\subsection{Structure of the Paper}
		The system model is introduced in Section \ref{sec:sysmodel}, which includes the network model, the communication model and the opinion evolution process. Section \ref{sec:prblmformul} presents the problem formulation and an illustrative example where the problem is solved in closed form. Then to extend the ideas of toy model to larger networks, the system model is represented as a dynamic Bayesian network. Centralized and decentralized algorithms are developed in Section \ref{sec:centraliz} and Section \ref{sec:decentraliz}, respectively. Complexity analysis of the algorithms is presented in Section \ref{sec:complexity}. Simulation results are discussed in Section \ref{sec:simres}. Finally, Section \ref{sec:conclus} concludes our work.
		
		\section{System Model}
		\label{sec:sysmodel}
		\subsection{Network Model}
		\noindent Consider an undirected graph $G = (V, E)$, where $V$ and $E$ are the set of vertices and edges, respectively. The vertex set $V$ is partitioned into two disjoint subsets $V = V_S \cupdot V_R$, where $V_S$ and $V_R$ denote the set of source nodes and regular nodes, respectively. The source nodes in-turn consist of smart sources $\Tilde{V}$ and random sources $V_r$, i.e., $V_S = \Tilde{V} \cupdot V_r$, that employ smart and random information spreading processes, respectively. Throughout the paper, we consider only one smart source node, i.e., $|\Tilde{V}| = |\{\Tilde{v}\}| = 1$, and one or more random source nodes in the network ($|V_r| \geq 1$). Without loss of generality, it is assumed that every source node $v_S \in V_S$ injects its own distinct messages into the network at each time slot. The regular nodes $v_R \in V_R$ facilitate the propagation of information across the network by forwarding the message to their neighbors. Each regular node $v_R \in V_R$ has a time-varying feed $F^{(v_R)}_t$ of a fixed finite size $L$. Upon reception of a message, it is stored in the feed in a FIFO (first-in-first-out) manner. Let $M$ be the set of messages circulated in the network, and let $U_{\Theta} = \Theta \cupdot \Bar{\theta}$ be the set of distinct message classes. The set $\Theta$ can be interpreted as a set of categories of competing products advertised by a company or ideologies in political campaigns, and each message is analogous to a specific advertisement or a particular propaganda, respectively, while class $\Bar{\theta}$ includes personal messages. We define the \textit{inclination} as the mapping $\mathcal{I}: M \rightarrow{} U_\Theta: m \mapsto \vartheta$, which maps each message to a class. The regular nodes can transmit messages of any class in $U_{\Theta}$, while each source node transmits messages corresponding to a fixed class in $\Theta$. Henceforth, it is assumed that the smart source injects messages of class $\Tilde{\theta} \in \Theta$ into the network. Next, we provide concrete definition of opinion, parameters governing beliefs and the overall strength of opinions of a user in social network.
		\begin{definition}
			Let $\{\alpha^{(v)}_{\theta, t}\}$ be the set of belief parameters that represent the affinity of node $v$ towards class $\theta$ at time $t \in \{0, 1, ..., T\}$, $\forall v \in V$ and $\forall \theta \in \Theta$. Then, for node $v$ the overall strength of opinions is defined as $\rho^{(v)}_t \triangleq \sum_{\theta \in \Theta}^{} \alpha_{\theta,t}^{(v)}$, and the opinion of node $v$ about the class $\theta$ is defined as $\mu^{(v)}_{\theta, t} \triangleq \alpha_{\theta, t}^{(v)}/\rho_t^{(v)}$. Consequently, the support of individual opinions is the simplex $[\mu^{(v)}_{\theta, t}]_{\theta \in \Theta} \in \mathbbm{R}^{|\Theta|}$, where $\sum_{\theta \in \Theta}^{} \mu^{(v)}_{\theta, t} = 1$. \label{def:opin}
		\end{definition} Next, we present the communication model, in which the interaction between the nodes is governed by their current opinions.
		
		\begin{figure*}
			\centering
			\includegraphics[width=0.95\textwidth]{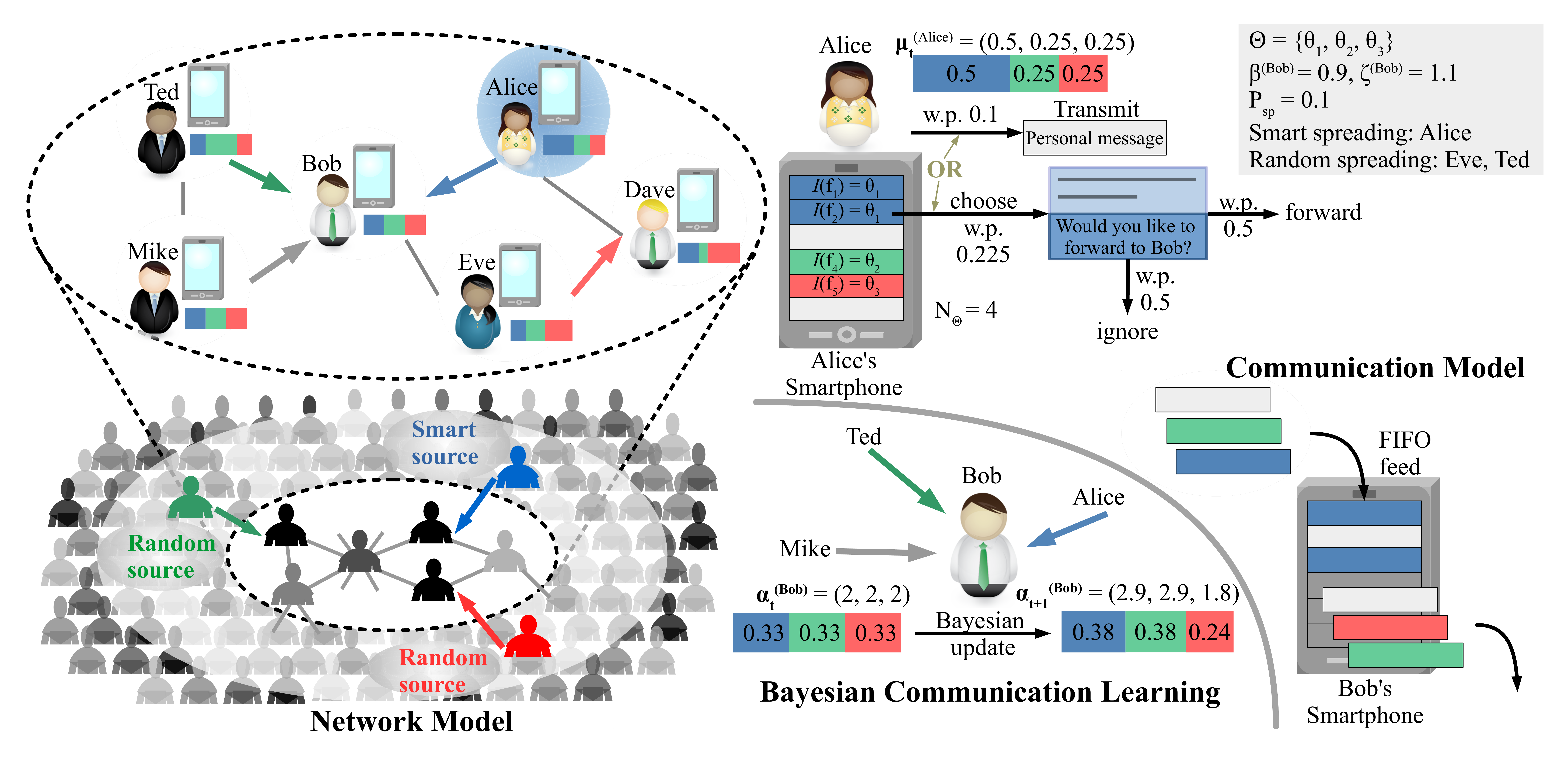}
			\caption{Illustration of the system model: Network structure, communication and evolution of opinions in a social network.}
			\label{fig:system_model}
		\end{figure*}
		
		\subsection{Communication Model}
		In this work, we adapt the broadcast-based communication model of \cite{sreenivasan2017information} to the synchronous gossip (push only). In applications like peer-to-peer advertisement dissemination and political campaigning, the communication is predominantly push-based since the target individuals are unaware of the impending message. In our model, a \textit{regular node} exhibits the following characteristics: 
		\begin{inparaenum}
			\item spontaneously generate and push personal messages (class $\Bar{\theta}$) with probability (w.p.) $P_{sp}$, or else (w.p. 1-$P_{sp}$)
			\item push (forward) a message $f$ of class $\Theta$ from its feed w.p. $P_f$ to one of its neighboring nodes.
		\end{inparaenum}
		In the latter, we assume that the message $f$ is chosen uniformly at random (u.a.r.) from the messages of class $\Theta$ in the feed and transmitted w.p. $\mu_\theta$, where $\theta = \mathcal{I}(f)$. Therefore, $P_f = \frac{\mu_\theta}{N_{\Theta}}$, where $N_\Theta$ is the number of messages of class $\Theta$ in the feed; the dependency of $P_{f}$ on the opinion $\mu_\theta$ is a reasonable choice, since users in social networks mostly forward messages that align with their opinion. We assume that the personal messages are not forwarded. A \textit{source } generates messages at rate $R_m$ and pushes them to one of its neighbors at every time step. In our model, we assume that at every node the messages corresponding to random sources (class $\Theta \setminus \Tilde{\theta}$) are forwarded to one of its neighbors u.a.r., while the messages of the smart source (the messages of class $\Tilde{\theta}$) are forwarded using certain mechanism (discussed in subsequent sections).
		
		\label{sec:commodel}
		
		Next, we describe the opinion evolution as a time-varying Dirichlet distribution. For concreteness, we introduce the Dirichlet distribution as follows.
		\begin{definition}
			The \textit{Dirichlet distribution} \cite{sudderth2006graphical} is defined as:
			\begin{equation}
			Dir(\alpha_1,...,\alpha_M) \triangleq  \frac{\prod_{i=1}^{M}\Gamma(\alpha_i)}{\Gamma\left(\sum_{i=1}^{M}\alpha_i\right)}\prod_{i=1}^{M}x_i^{\alpha_i-1},
			\end{equation}
			where $[x_i]_{1\leq i \leq M}$ is the support with $\sum_{i=1}^{M} x_i= 1$, where $x_i > 0$, $\forall i \in \{1, 2, ..., M\}$, and $\{\alpha_i,...,\alpha_{M}\}$ are the parameters of the distribution.
		\end{definition}
		
		\subsection{Evolution of Opinions}
		\label{sec:evolopin}
		Similar to \cite{azzimonti2018social}, the individuals in the social network are modeled as Bayesian learning agents. These agents update their beliefs upon receiving messages of class $\Theta$, while the personal messages do not modify their beliefs. We extend the model in \cite{azzimonti2018social} to a multi-polar society, where more than two classes of messages are propagated. Therefore, the prior belief of each node is modeled as a Dirichlet distribution using the belief parameters described in Definition \ref{def:opin}. With every incoming message, the corresponding belief parameter is updated based on inclination of the message. Since a node can receive messages from multiple neighbors, we treat the incoming messages corresponding to multiple classes as multinomial observations. Consequently, the posterior belief is in-turn a Dirichlet distribution, since it forms a conjugate pair with multinomial distribution.
		
		\subsubsection{Bayesian Communication Learning}
		We assume that the nodes have heterogeneous learning behaviors, where certain nodes trust the incoming messages strongly and make a significant update to their belief, while others are \textit{stubborn} nodes that must be persuaded more to alter their beliefs. Moreover, some nodes have higher retention of social learning than others. For every node $v \in V_R$, the two aforementioned behaviors are captured using the node specific parameters, $\zeta^{(v)}>0$  and $\beta^{(v)} \in ]0, 1[$, respectively. Let $n^{(v)}_{\theta, t}$ be the number of incoming messages of class $\theta \in \Theta$ at time $t$. Node $v$ updates its belief parameter as per the following rule:
		\begin{equation}
		\alpha^{(v)}_{\theta, t} = \beta^{(v)} \alpha^{(v)}_{\theta, t-1} + \zeta^{(v)} n^{(v)}_{\theta, t}.
		\label{eq:blfupdate1}
		\end{equation}
		Therefore, the posterior belief is given by:
		\begin{equation}
		\begin{split}
		(G^{(v)}_{t} &(\theta_1)...G^{(v)}_{t}(\theta_{|
			\Theta|}))|\mathbf{n}^{(v)}_1 ,...,\mathbf{n}^{(v)}_{t} 
		\\
		\sim  Dir&(\beta^{(v)} \alpha^{(v)}_{1,t-1} + \zeta^{(v)} n^{(v)}_{1, t},...,\beta^{(v)} \alpha^{(v)}_{|\Theta|,t-1} + \zeta^{(v)} n^{(v)}_{\theta, t}), \label{eq:dirichletPredict}
		\end{split}
		\end{equation}	
		where $G_t$ is a random distribution over $\Theta$ at time $t$, and $\mathbf{n}^{(v)}_{\tau} = \left(n^{(v)}_{\theta, \tau}\right)_{\theta \in \Theta}$. It can be observed in Eqs. \eqref{eq:blfupdate1}-\eqref{eq:dirichletPredict}, that the magnitude of belief parameter (hence the opinion) is proportional to the number of incoming messages of the corresponding class.
		
		The entire system model and the opinion dynamics is illustrated in Fig. \ref{fig:system_model}. There are three sources which transmit messages of distinct classes $\theta_1$, $\theta_2$ and $\theta_3$, where class $\theta_1$ messages are generated by the smart source. For illustration, we zoom-in to observe a sub-network of 6 users at time $t$. Alice chooses a message $f_2$ from her feed w.p. $\frac{1-P_{sp}}{N_{\Theta}} = \frac{0.9}{4} = 0.225$. which consists of a recommendation to forward the message to Bob. Then, Alice forwards the message to Bob w.p. $\mu^{(Alice)}_{\theta_1, t} = 0.5$. Note that the forwarding recommendation is smart since Bob is more influential (clarified later) than her neighbor Dave. On the other hand, Eve selects a message of class $\theta_3$ from her feed which consists of a random recommendation to forward the message to Dave. Ted and Mike push message (u.a.r.) of class $\theta_2$ and a personal message to Bob, respectively. Also, the FIFO feed and belief parameter update by Bob upon receiving messages from Ted, Alice and Mike is depicted in the Fig. \ref{fig:system_model}. Note that the personal message from Mike does not modify the opinion of Bob. The goal of the smart source is to generate smart forwarding recommendations for those nodes that have chosen message of class $\Tilde{\theta}$ (in this case Alice) from their respective feeds so that the overall opinion towards the smart source is maximized. In the next section, we provide a mathematical treatment of the opinion maximization problem.
		
		\section{Problem Formulation}
		\label{sec:prblmformul}
		\noindent In this section, first a formal definition of the opinion maximization problem is provided, which is followed by the discussion of a toy example where we solve the problem in closed form. Then, in order to extend the ideas for general case (arbitrary connected networks and considering the factor of time), we present dynamic Bayesian network and influence diagrams, which will be used subsequently to develop algorithms to achieve opinion maximization. To begin, we define the \textit{action} taken by the nodes in the network.
			
		\begin{definition} \textit{Action of node $v$:}
		Let node $v$ choose one of the messages from its feed, and let $w \in N(v)$ be the forwarding recommendation associated with that message at time $t$. Then the action of node $v$ is $a^{(v)}_t = w$.
		\end{definition} 
		We denote the joint action as $\mathbf{a}_t = \left( a^{(v)}_t \right)_{v \in \Tilde{\mathcal{V}}_t}$, where the set $\Tilde{\mathcal{V}}_t$ denotes the set of nodes that have chosen message of class $\Tilde{\theta}$ from their feeds at time $t$. The objective of the smart source is to maximize the opinion of individuals in the social network towards class $\Tilde{\theta}$. The formal definition of opinion maximization is given as follows:
		\begin{definition}
			Let $(\mathbf{a}_t)_{0\leq t \leq T-1}$ be the sequence of joint actions and $\bm{\alpha}_0=[\alpha^{(v)}_{\theta,0}]_{\substack{v \in V \\ \theta \in \Theta}}$ be belief parameters at time $t=0$. The objective of the smart source is to maximize the expected total opinion of class $\Tilde{\theta}$ at time $T$, which is given as:
			\begin{equation}
			\underset{\left(\mathbf{a}_t\right)_{0\leq t\leq T-1}}{\text{maximize}} \hspace{2mm} \mathbb{E}\left\{ \sum\limits_{v \in V} \mu^{(v)}_{\Tilde{\theta},T} \mid  \left(\mathbf{a}_t\right)_{0\leq t\leq T-1}, \bm{\alpha}_0 \right\}.
			\label{eqn:optimprob1}
			\end{equation}
			\label{def:opinmax}
		\end{definition}
		In Appendix \ref{ap:prop1}, the opinion maximization problem is analyzed for a simplified version of the communication model described in \ref{sec:commodel}. In view of Eq.~\eqref{eq:prop1last} in Appendix \ref{ap:prop1}, we define the reward as follows:	
		\begin{definition}
			The \textit{reward} obtained by node $v$ when it pushes a message of class $\Tilde{\theta}$ to $w \in N(v)$ is defined as: $r^{(v)}(w) \triangleq \frac{\mu^{(w)}_{\Tilde{\theta}, t} (1-\mu^{(w)}_{\Tilde{\theta}, t})}{\mu_{\Tilde{\theta}, t}^{(w)} + \alpha_{\Tilde{\theta}, t}^{(w)} \beta^{(w)}/\zeta^{(w)}}$.
			\label{def:rwd}
		\end{definition}
		
		\begin{observation}
			\label{rmk1}
			In Definition \ref{def:rwd}, the reward indicates the change in opinion of node $w$ about class $\Tilde{\theta}$. Moreover, the best myopic action is to push the message to a node with weakly neutral opinion. For instance, let $\beta, \zeta = 1$ and $\alpha_{\theta, 0} = 1$, $\forall \theta \in \Theta$, then the instantaneous reward $r_t \approx \frac{\mu_{\Tilde{\theta},t} (1-\mu_{\Tilde{\theta},t})}{(\mu_{\Tilde{\theta},t}+\alpha_{\Tilde{\theta},t})}$ is maximized when $\mu_{\Tilde{\theta},t} \approx 1/2$ (neutral) and $\alpha_{\Tilde{\theta},t} \downarrow 1$ (weak).
		\end{observation}
		
		\subsection{A Toy Example of Opinion Maximization}
		\label{sec:toyex}
		In this section, we present a toy model, where the problem is solved in closed form. The purpose of the toy model is to obtain some insights, which are later used to develop algorithms for large-scale networks.  The toy model consists of 4 nodes with 2 transmitters ($x$ and $y$) which can push message of class $\theta_1 \in \Theta$, to one of 2 receivers ($c$ and $d$), independently. The actions of the nodes $x$ and $y$ are denoted as $a^{(x)} \in \{c, d\}$ and $a^{(y)}  \in \{c, d\}$, respectively\footnote{In this section, subscript $t$ is omitted}. Moreover, the opinions of nodes $x$ and $y$ about class $\theta_1$ is considered to be $\mu^{(x)}_{\theta_1} = \mu^{(y)}_{\theta_1} = 1$; therefore the action $a^{(i)} = j$ is equivalent to saying node $i$ pushes message to node $j$, $\forall i \in \{x,y\}$ and $\forall j \in \{c,d\}$. Both the receivers are Bayesian learning agents with the opinions about the two message classes ($\theta_1$ and $\theta_2$) denoted as $\mu^{(i)}_{\theta_1}$ and $\mu^{(i)}_{\theta_2}$ ($i \in \{c, d\}$), respectively.          
		Both the nodes $x$ and $y$ are oblivious to the actions of the other. If only node $x$ pushes to node $c$, then from Eq.~\eqref{eq:prop1last} the change in opinion is given as: $\mu^{(c)}_{\theta_1} \overline{\mu^{(c)}_{\theta_1}} (\mu^{(c)}_{\theta_1} +\alpha^{(c)}_{\theta_1} \beta^{(c)}/\zeta^{(c)})^{-1} = \frac{\alpha^{(c)}_{\theta_2} \zeta^{(c)}}{(\beta^{(c)} \rho^{(c)} + \zeta^{(c)}) \rho^{(c)}}$, which we call the individual reward $r_c$. Similarly, the individual reward $r_d = \frac{\alpha^{(d)}_{\theta_2} \zeta^{(d)}}{(\beta^{(d)} \rho^{(d)} + \zeta^{(d)}) \rho^{(d)}}$. The nodes $x$ and $y$ have the knowledge of opinions of the nodes $c$ and $d$. Consequently, they know the individual rewards $r_c$ and $r_d$. We call the joint reward as the total change in opinion when both the nodes $x$ and $y$ transmit to nodes $c$ or $d$. The joint rewards for all the combinations of joint actions (of $x$ and $y$) are given in Table \ref{tbl:toyrwd}, and are not known to $x$ and $y$ a priori. The joint rewards are given by: 
		\begin{equation}
		R_{cd} = \sum_{j \in \{c,d\}}^{}\frac{\alpha^{(j)}_{\theta_2} \zeta^{(j)}}{(\beta^{(j)} \rho^{(j)} + \zeta^{(j)}) \rho^{(j)}},
		\end{equation}
		and
		\begin{equation}
		R_{ii} = \frac{2 \alpha^{(i)}_{\theta_2} \zeta^{(i)}}{(\beta^{(i)} \rho^{(i)} + 2 \zeta^{(i)}) \rho^{(i)}}, \hspace{2mm} \forall i \in \{c,d\}.
		\end{equation}
		The objective is to determine the strategy, i.e., the best choice of actions, such that the total opinion ($\mu^{(c)}_{\theta_1}+\mu^{(d)}_{\theta_2}$) is maximized. In this scenario, we state the following proposition that provides the condition where pure strategy (rule that maps individual rewards to the actions) does not yield the maximum reward, and also determine the mixed strategy (probability distribution over actions/pure strategies) that results in the maximum reward.
		\begin{table}[h]
			\begin{minipage}[b]{0.5\linewidth}
				\centering
				\begin{tabular}{cc|c|c|}
					\multirow{4}{0mm}{\vspace{7mm} $y$} & \multicolumn{3}{c}{\hspace{6mm} $x$} \\
					\multicolumn{1}{c}{} & \multicolumn{1}{c}{} & \multicolumn{1}{c}{$c$} & \multicolumn{1}{c}{$d$} \\
					\cline{3-4}
					& & & \\[0pt]
					& $c$ & $R_{cc}$ & $R_{cd}$  \\[5pt]
					\cline{3-4}
					& & & \\[0pt]
					& $d$ & $R_{cd}$ & $R_{dd}$   \\[5pt]
					\cline{3-4}
					\\ [5pt]
				\end{tabular}
				\vspace{-3mm}
				\caption{Reward Table.}
				\label{tbl:toyrwd}
			\end{minipage}
			\begin{minipage}[b]{0.45\linewidth}
				\centering
				\includegraphics[width=40mm, height=20mm]{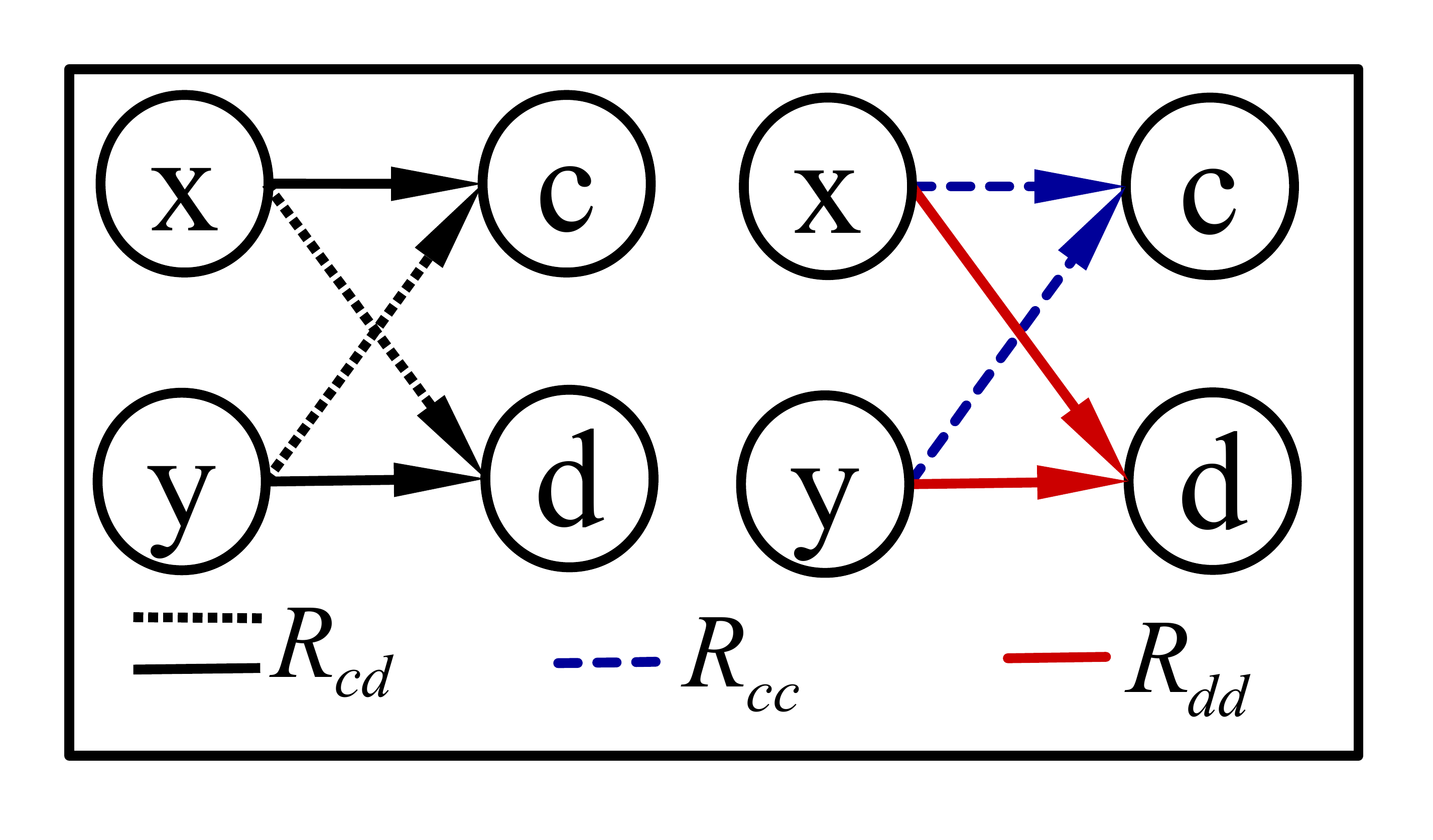}
				\captionof{figure}{Toy Example.}
				\label{fig:toyex}
			\end{minipage}
		\end{table}
		\begin{proposition}
			In the aforementioned \textit{toy example}, if 
			$\frac{1}{1+\eta^{(c)}} < \frac{r_d}{r_c} < \min\left\{ 1, \left(\frac{1+\eta^{(c)}}{1+2\eta^{(c)}} \right) \left( \frac{1+2 \eta^{(d)}}{1+\eta^{(d)}}\right) \right\}$, then the maximum reward is given by the mixed strategy, $\pi = (p, \Bar{p} = 1-p)$, where $p = P(a^{(x)} = c) = P(a^{(y)} = c) = \left(1+\frac{R_{cd} -R_{cc}}{R_{cd} + R_{dd}}\right)^{-1}$ and $\eta^{(i)} = \frac{\zeta^{(i)}}{\beta^{(i)} \rho^{(i)}}$, $\forall i \in \{c,d\}$.
			\label{prop:toymodel}
		\end{proposition}
		\textit{Proof}: See Appendix \ref{app:toymodelproof}. 
		
\begin{figure}[h]
			\minipage{0.24\textwidth}
			\includegraphics[width=0.9\textwidth]{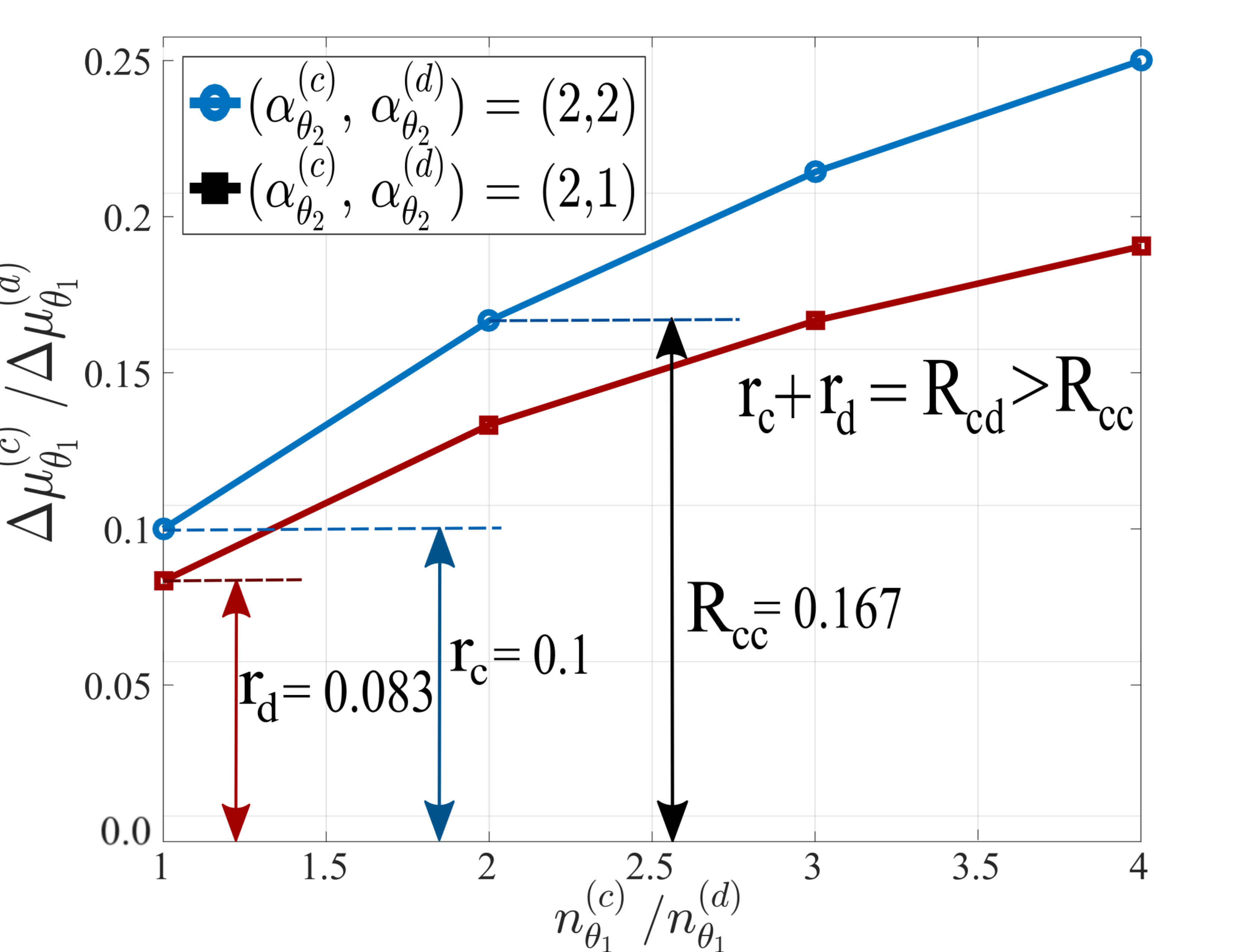}
			\caption{Diminishing returns.\\ \hspace{1mm}}
			\label{fig:diminishing_utility2}
			\endminipage
			\minipage{0.24\textwidth}
			\includegraphics[width=0.9\textwidth, height=0.7\textwidth]{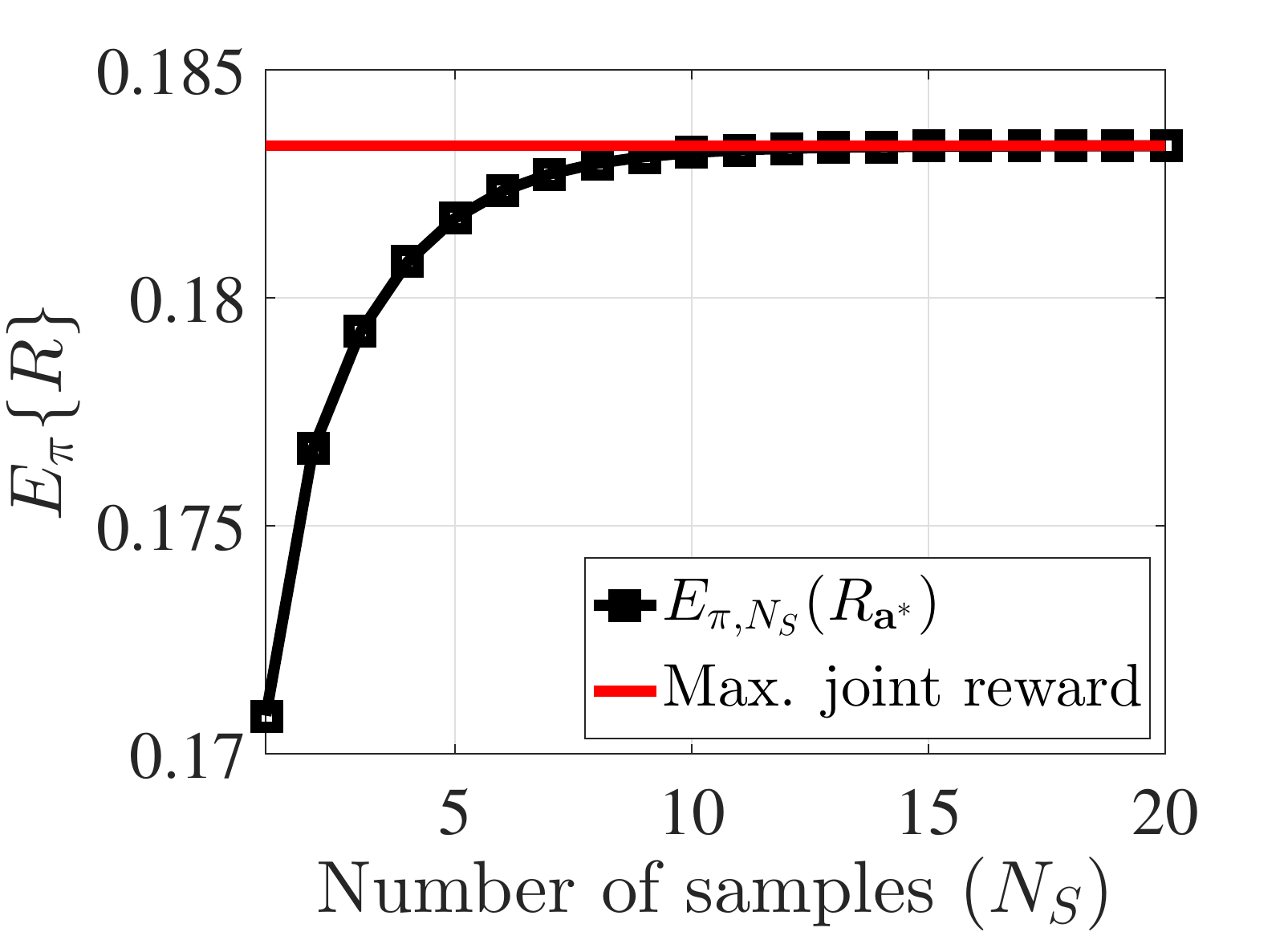}
			\caption{Convergence of expected reward to the optimal value.}
			\label{fig:toyconvg}
			\endminipage
		\end{figure}

		Informally speaking, the proposition states that in the toy example if the nodes $c$ and $d$ have similar beliefs, then the better strategy for nodes $x$ and $y$ is to take distinct actions (non-diagonal elements in Table \ref{tbl:toyrwd}). This phenomenon is illustrated using a numerical example in Fig. \ref{fig:diminishing_utility2}. It can be observed that the reward $\Delta \mu_{\theta_1}$ exhibits diminishing returns with respect to the number of incoming messages $n_{\theta_1}$, which underscores the fact that taking distinct actions yields higher reward $R_{cd} = r_c + r_d$. In large-scale social networks, many nodes have common neighbors (e.g., mutual friends).  Taking selfish actions implies that multiple nodes try to persuade one of their common neighbors to change its opinion. This could lead to lower joint rewards and also restrict information spreading. On the other hand, using mixed strategy helps in better information spreading and yields better joint rewards.
		
		\begin{definition}
			Let $h(1)$ and $h(2)$ be the rewards obtained by taking two distinct actions (with $\mathbf{h} = \left(h(1), h(2)\right)$). Then for $i \in \{1,2\}$, the Boltzmann distribution (also called soft-max) is defined as:
			\begin{equation}
			\Xi(\mathbf{h}, i, \mathcal{T}) \triangleq \exp(h(i)/\mathcal{T})/\left(\Sigma_{j=1}^{2}{\exp(h(j)/\mathcal{T})}\right),
			\end{equation}
		\end{definition}
		where $\mathcal{T}>0$ is the temperature parameter.
		\begin{remark}
			\label{remarkBoltz}
			Given a mixed strategy $\pi = (p, \Bar{p})$, if $p>\Bar{p}$ and $h(1) > h(2) > 0$, then $\exists$ $\mathcal{T} > 0$ such that $(p, \Bar{p}) = \left(\Xi(\mathbf{h}, 1, \mathcal{T}), \Xi(\mathbf{h}, 2, \mathcal{T})\right)$ (the proof is straightforward). This implies that in the aforementioned toy example with two actions, the mixed strategy can be obtained exactly using the individual rewards and Boltzmann distribution by tuning $\mathcal{T}$ appropriately. However, if the number of actions is greater than 2, then it can be easily verified that the mixed strategy can only be approximately obtained.
		\end{remark}
		
		\begin{remark} \textit{Sampling Improves Expected Reward:}
			\label{rmk:samp}
			Consider a central controller which is capable of controlling joint actions $\mathbf{a} = (a^{(x)}, a^{(y)})$ and observe the joint rewards. It can try multiple joint actions offline and determine the best joint action (that gives the maximum joint reward) from history, following which it is executed. In particular, the central controller samples joint actions from the distribution $\bm{\pi} = \pi\times \pi$, $N_S$ times (denoted as $\mathbf{a} \overset{\scriptsize{N_S}}{\sim} \bm{\pi}$). The best joint action is given by: $\mathbf{a}^* = \text{argmax}_{\mathbf{a} \overset{\scriptsize{N_S}}{\sim} \bm{\pi}} R_{\mathbf{a}}$. Let $p^\prime = 1-2p\Bar{p}$ and $p^{\prime \prime} = (1-p^{2}/(p^{2}+\Bar{p}^{2}))$. The expected joint reward is given by: 
			\begin{equation}
			\begin{split}
			\mathbbm{E}_{\bm{\pi}}(R_{\mathbf{a^*}}) &= \left(1-{p^{\prime}} ^{N_s}\right) R_{cd} \\
			&+ p^{\prime N_s} \left( (1-{p^{\prime \prime}})^{N_s} R_{cc} + p^{\prime \prime N_s} R_{dd} \right).
			\end{split}
			\end{equation}
			As shown in Fig. \ref{fig:toyconvg}, we can observe that $\lim\limits_{N_s \rightarrow \infty} \mathbbm{E}_{\bm{\pi}}(R_{\mathbf{a^*}}) = R_{cd}$, which is the maximum joint reward in Table \ref{tbl:toyrwd}.
		\end{remark}
		
		The toy model provides three crucial insights: \begin{inparaenum} \item mixed strategy yields better reward than selfish actions (Proposition~\ref{prop:toymodel}), \item the individual rewards and Boltzmann distribution can be used to obtain the mixed strategy (Remark \ref{remarkBoltz}), and \item sampling improves the expected reward (Remark \ref{rmk:samp}). \end{inparaenum} In the toy model, we considered a single snapshot of a small network. We extend the ideas of the toy model to larger networks  and also take the factor of time into account. Therefore, we begin with the DBN representation of opinion dynamics as follows.
		
		\subsection{Representation of Opinion Evolution in the Network as a Dynamic Bayesian Network}
		
		Dynamic Bayesian networks (DBNs) are probabilistic graphical models where the nodes stand for random variables, and their conditional dependencies and temporal relationships are represented through a directed acyclic graph \cite{koller2009probabilistic}. Two time-slices are required to fully represent the dynamics: one for indicating conditional relationship between random variables, and the other for depicting the causal dependence. This representation helps in developing approximate iterative algorithms for sequential decision problems.
		
		Let $X^{(v)}_{\theta, t}$ be a random variable that represents the belief parameter $\alpha^{(v)}_{\theta, t}$, where $v \in V$ and $\theta \in \Theta$.  We construct a random matrix $X_t = [X^{(v)}_{\theta,t}]_{\substack{v \in V \\ \theta \in \Theta}}$, which captures the belief parameters of the entire network at time $t$. Similarly, let $\Omega_t = [\Omega^{(v)}_t]_{v \in V}$ be a random vector, where each element $\Omega^{(v)}_t$ is a random variable representing the inclination of message $m^{(v)}_t$ chosen by node $v$ from its feed at time $t$. Let $F_t = [F^{(v)}_{j,t}]_{\substack{v \in V \\ 1 \leq j \leq L}}$ be a random matrix where each element is a random variable that represents the $j^{\text{th}}$ message in the feed of node $v$ at time $t$. Finally, let $A_t = [A^{(v)}_t]_{v \in V}$ be a random vector, where $A^{(v)}_t$ represents the action of node $v$ at time $t$. In view of the communication model and opinion evolution model described in \ref{sec:commodel} and \ref{sec:evolopin}, the overall opinion dynamics in the network can be explained as follows: Nodes choose messages from their feeds, and decide whether the message should be forwarded or not based on their current opinions. Then, based on the actions of the nodes and the class of the chosen messages, the beliefs of the recipient nodes are updated. The newly received messages update the feeds by occupying the top positions, while pushing out the older messages. We represent these dynamics using a DBN as shown in Fig. \ref{fig:qopin_dbn}.
		
		\subsection{From DBN to Influence Diagram}	
		\label{sec:infludiag}
\begin{figure*}[t]
			\minipage{2cm}
			\includegraphics[width=2cm,height=3cm]{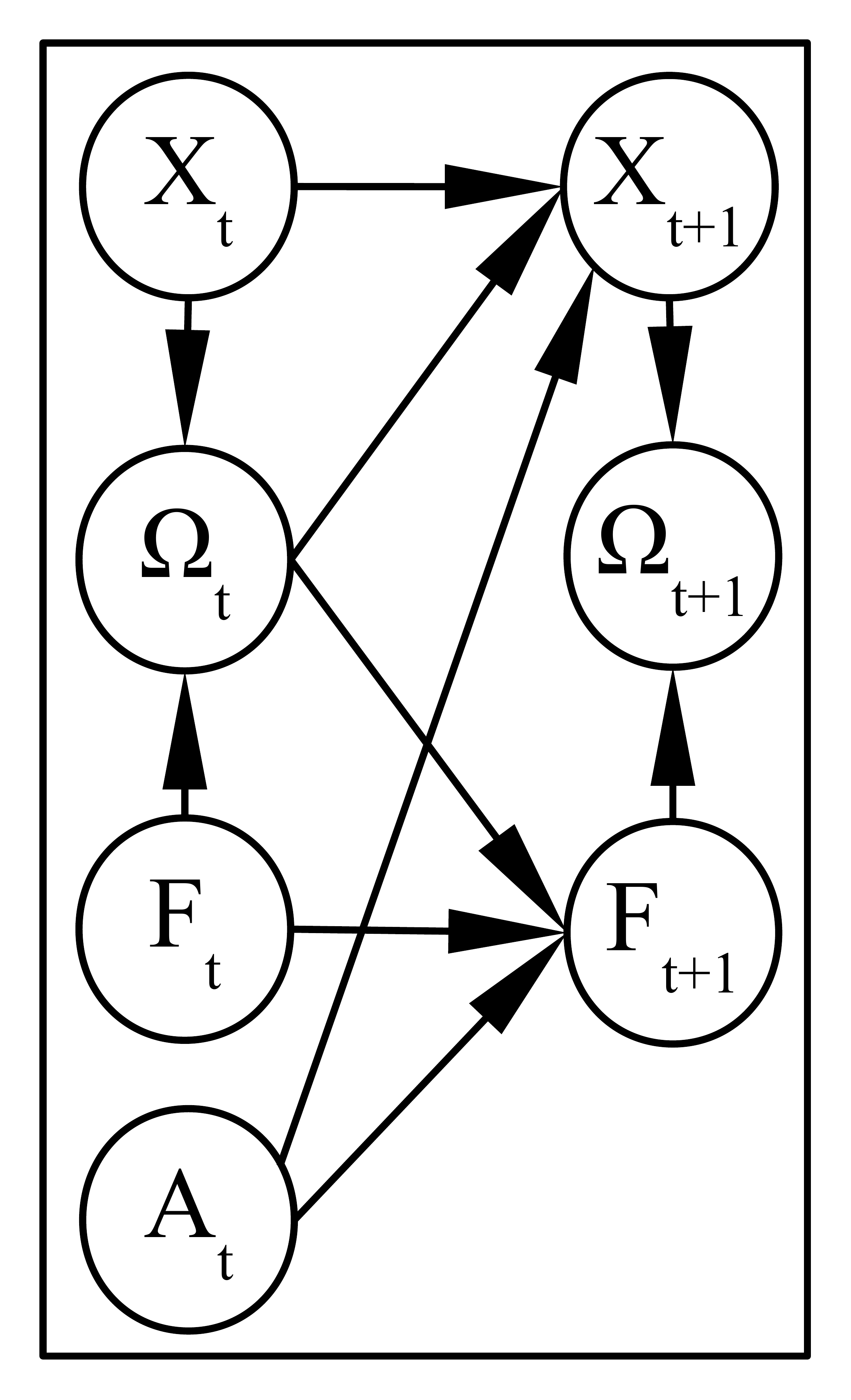}
			\caption{DBN representation.}
			\label{fig:qopin_dbn}
			\endminipage
			\quad
			\minipage{5.1cm}
			\includegraphics[width=5.3cm,height=3.4cm]{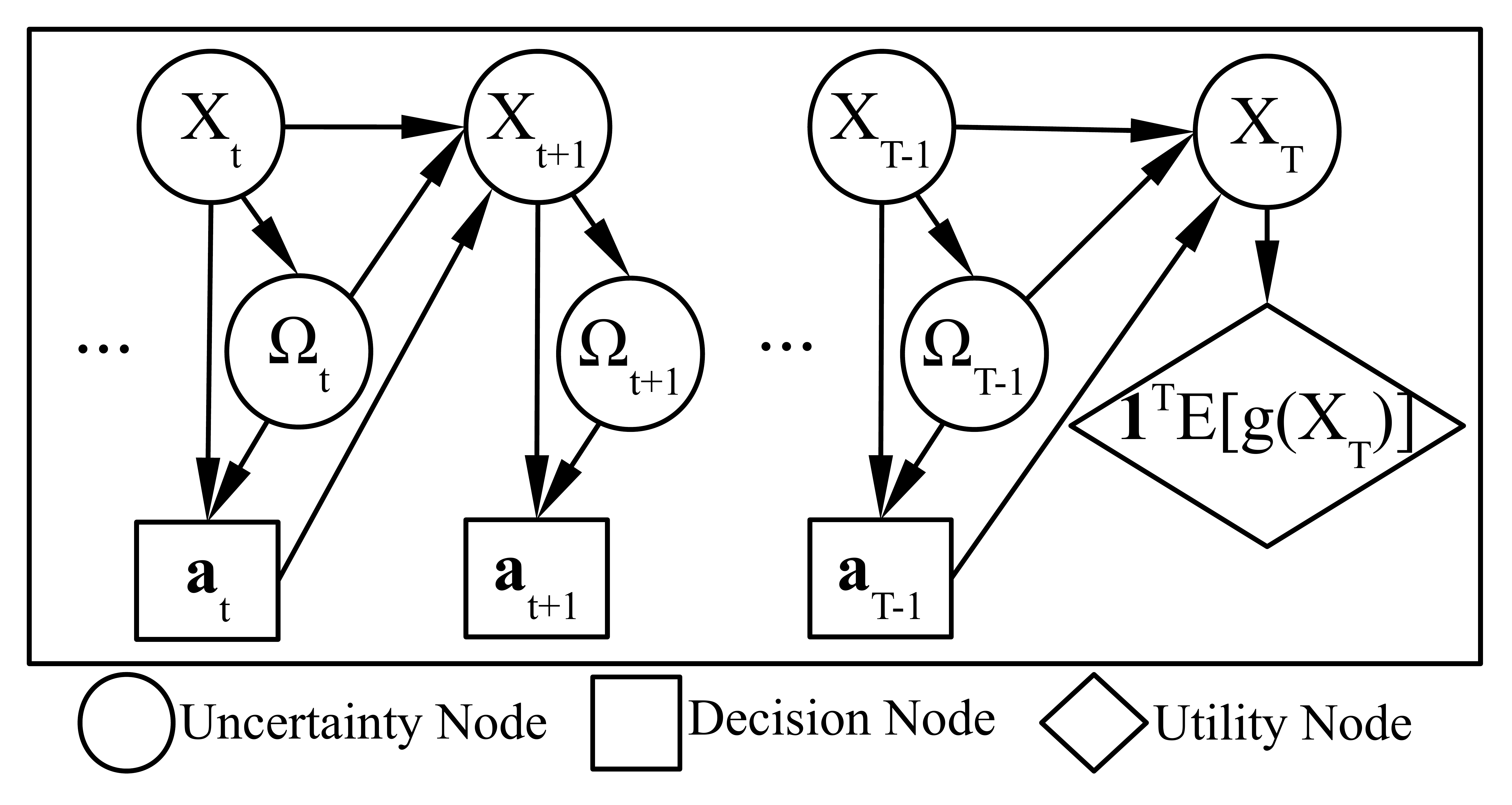}
			\caption{Influence diagram.}
			\label{fig:infludiag}
			\endminipage
			\quad
			\minipage{9.9cm}
			\includegraphics[width=10cm,height=3.4cm]{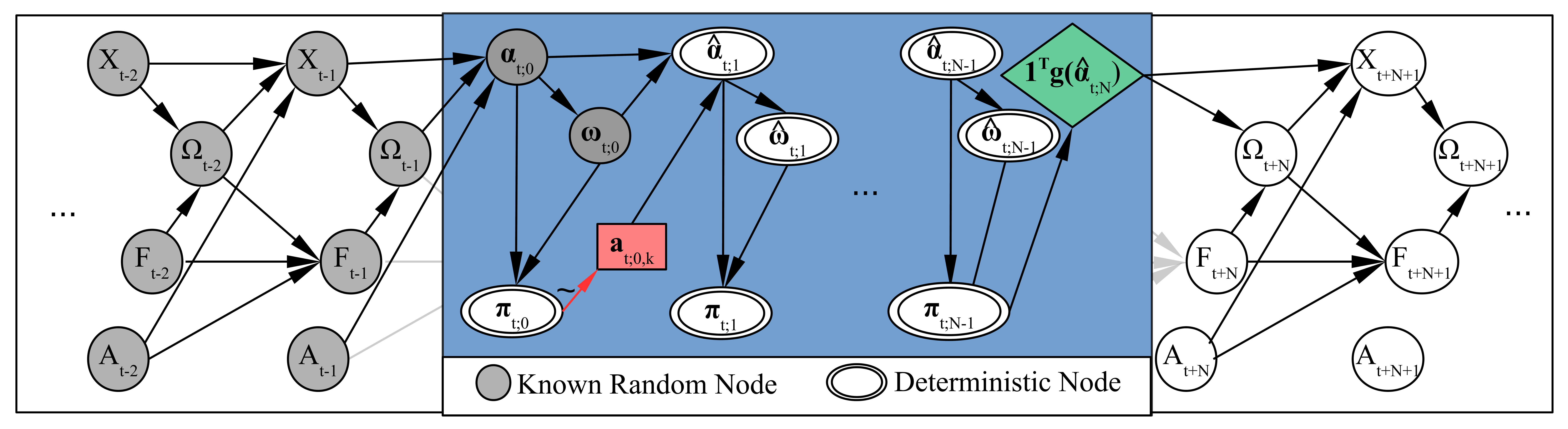}
			\caption{N-step look-ahead algorithm.}
			\label{fig:qopin_dbn_cen}
			\endminipage
			\vspace{-5mm}
		\end{figure*}

		In decision theory, some variables of a DBN are converted to decision variables and utility variables, and the whole model is alternatively called an influence diagram. In our model, the influence diagram (Fig. \ref{fig:infludiag}) is constructed from DBN as follows. We assume that $F_t$ cannot be observed; hence, the uncertainty node $F_t$ of the DBN and all the associated edges (both incoming and outgoing) are removed from the influence diagram. Even though the removal is not optimal, it makes our analysis tractable. The uncertainty node $A_t$ is converted to a decision node $\mathbf{a}_t$\footnote{Henceforth, unless stated otherwise, the outcomes of random variables are indicated by lower case bold letters.}. Before determining action $\mathbf{a}_t$, the random variables $X_t$ and $\Omega_t$ are observed. Hence, informational arcs are connected from the observable nodes to the decision node $\mathbf{a}_t$. The goal of the problem is to determine the optimal sequence of actions $\mathbf{a}_0,...,\mathbf{a}_{T-1}$, such that the total expected opinion as described in Definition \ref{def:opinmax} is maximized (the rigorous mathematical treatment is given in Section \ref{sec:centraliz}). The influence diagrams and solving decision problems are discussed comprehensively in \cite{koller2009probabilistic}. \section{Centralized Algorithms}
		
		\label{sec:centraliz}
		\noindent 	
		In this section, starting from the optimization problem in Definition \ref{def:opinmax}, we use the influence diagram and ideas from the toy model (Section \ref{sec:toyex}) to construct a framework, using which we develop centralized iterative algorithms. The central controller possesses $\forall v \in V$ and $\forall \theta \in \Theta$ the knowledge of the opinion $\mu^{(v)}_{\theta, t}$, the overall strength $\rho^{(v)}_t = \sum_{\theta}^{} \alpha^{(v)}_{\theta, t}$, the probability of spontaneous transmission ${{P_{sp}}}$, the global topology of the network $G$, and the locations of all the source nodes (the quantities $\mu^{(v)}_{\theta, t}$ and $\rho^{(v)}_t$ can be estimated from the users reviews/ratings, and $\beta^{(v)}$, $\zeta^{(v)}$ and $P_{sp}$ can be estimated from the history of messages sent by the users\footnote{Estimating $\mu^{(v)}_{\theta, t}$, $\rho^{(v)}_t$, $\beta^{(v)}$, $\zeta^{(v)}$ and $P_{sp}$ are beyond the scope of this paper.}). Let $g: \bm{\alpha}=[\alpha^{(v)}_{\theta}]_{\substack{v \in V \\ \theta \in \Theta}} \rightarrow \bm{\mu}_{\Tilde{\theta}} = [\alpha^{(v)}_{\Tilde{\theta}}/\sum\limits_{\theta \in \Theta} \alpha^{(v)}_\theta]_{v \in V}$ be the function that maps belief parameters to the opinions corresponding to class $\Tilde{\theta}$. Now, the objective function in Definition \eqref{def:opinmax} is alternatively given as: $\mathbbm{1}^T \mathbbm{E}[g(X_T) \mid  \left(\mathbf{a}_t\right)_{0\leq t\leq T-1}, \bm{\alpha}_0]$.
		From the influence diagram, it can be observed that given $X_t$, the action $\mathbf{a}_t$ does not depend on past observations. Moreover, $X_t$ is observed at every time step before action $\mathbf{a}_t$ is decided, which decreases the uncertainty in total opinion at time $T$.  Hence, instead of determining the sequence of actions $(\mathbf{a}_t)_{0 \leq t \leq T-1}$ upfront at time $t=0$, the action $\mathbf{a}_t$ that provides the maximum total opinion at time $T$ can be determined at every time $t$. Therefore, Eq.~\eqref{eqn:optimprob1} is modified to the following decision problem:
		\begin{equation}
		\mathbf{a}_t^* = \underset{\mathbf{a}_t}{\text{argmax}} \bm{\pi}^*_t \mid (\bm{\pi}^*_{\tau})_{t \leq \tau \leq T-1} = \hspace{-2mm} \underset{ (\bm{\pi}_{\tau})_{t \leq \tau \leq T-1}}{\text{argmax}} \hspace{-2mm} \mathbbm{1}^T \mathbbm{E}_{\psi} \{g(X_T) \mid \bm{\alpha}_{t}\}, 
		\label{eqn:optimprob2}
		\end{equation}
		where $\bm{\pi}_{\tau}$ is the probability distribution over joint actions at time $\tau$ with $\bm{\pi}_\tau(A_\tau) = P(A_{\tau} \mid \Omega_{\tau}, X_{\tau})$, $\mathbf{a}_t^*$ is an optimal action at time $t$, and $\psi = f(X_T \mid \bm{\alpha}_t) $ is obtained by marginalization as follows:
		\begin{flalign}
		\psi = \int\limits_{X_{t+1:T-1}}^{} \sum\limits_{\substack{\Omega_{t:T-1} \\A_{t:T-1}}}^{} \prod_{\tau = t}^{T-1} & f(X_{\tau+1} \mid X_\tau, \Omega_\tau, A_\tau) \notag  \\ 
		& P(\Omega_\tau \mid X_\tau) \bm{\pi}_{\tau}(A_\tau).
		\label{eq:psi}
		\end{flalign}
		To determine $\psi$ exactly, all the conditional densities in Eq.~\eqref{eq:psi} must be computed, which is computationally expensive (exponential computational complexity). This makes the problem in Eq.~\eqref{eqn:optimprob2} to be intractable. In order to address this issue, we construct the following framework.
		
		\subsection{A Framework for Centralized Algorithms}
		\label{sec:framework}
		The basic idea of the framework is to approximate $\psi$ to conveniently compute the objective function $\mathbbm{1}^T \mathbb{E}_{\psi}\{g(X_T \mid \bm{\alpha}_t)\}$.
		To be precise, we approximate $\psi$ by making the conditional distributions given in Eq.~\eqref{eq:psi} degenerate. This enables us to iteratively compute the objective function. In this regard, three approximations are provided, whose inherent assumptions are explained in \ref{sec:camo}.
		
		\textit{Approximation 1:}
		Obtaining the mixed strategy profile $\bm{\pi}_\tau$ is computationally demanding. Hence, we assume that the distribution of actions taken by the nodes are independent, resulting in the constraint (approximation):
		\begin{equation}
		\bm{\pi}_\tau(A_\tau) \approx \prod\vspace{-1mm}_{\vspace{-1mm}_{v \in V}}^{} \pi_\tau(A^{(v)}_{\tau}),
		\label{eq:indep}
		\end{equation}
		where $\pi_\tau(A^{(v)}_{\tau}) = P_\tau(A^{(v)}_{\tau} \mid \Omega_\tau, X_\tau)$ is the mixed strategy of node $v$. Given this approximation, we assume that if $\Omega_{\tau}$ and $X_{\tau}$ are observed, then we can obtain the joint distribution $\bm{\pi}_\tau$.
		
		Due to aforementioned approximations, the action $\mathbf{a}^*_t$ in Eq.~\eqref{eqn:optimprob2} is no more optimal. Therefore, using Remark \ref{rmk:samp},  we sample $\mathbf{a}_t$ from $\bm{\pi}_t$, and use $\left(\bm{\pi}_\tau\right)_{t+1 \leq \tau \leq T-1}$ to approximately compute the objective function. Moreover, for tractability the objective function is modified to $\mathbbm{1}^T g\left( \mathbb{E}_{\psi}\{X_T \mid \bm{\alpha}_t, \mathbf{a}_t\} \right)$, which results in the following decision problem:
		\begin{equation}
		\mathbf{a}_t^* = \underset{\mathbf{a}_t \overset{\scriptsize{N_S}}{\sim} \bm{\pi}_t}{\text{argmax}} \mathbbm{1}^T g\left( \mathbb{E}_{\psi}\{X_T \mid \bm{\alpha}_t, \mathbf{a}_t\} \right).
		\label{eqn:optimprob3}
		\end{equation}
		
		\textit{Approximation 2:} Given the belief parameters $X_{\tau}$, we assume that the maximum a posteriori (MAP) estimate $\hat{\bm{\omega}}_{\tau} = \text{argmax}_{\omega \in \Theta}P(\Omega_{\tau} = \omega \mid X_{\tau})$ can be obtained. Then we approximate the probability of choosing a message of class $\Theta$ to be degenerate around $\hat{\bm{\omega_\tau}}$. Also, note that probability of spontaneous transmission $P_{sp}$ can be interpreted as the probability of choosing message of class $\Bar{\theta}$. Moreover, every node in the network chooses message from its feed independently. Therefore, the conditional distribution of choosing message of any class in $U_{\Theta}$ is approximated as:
		\begin{equation}
		\begin{split}
		{P}(\Omega_{\tau} \mid X_{\tau}) \approx \prod\limits_{v \in V}^{} (1-P_{sp}) & \delta\left(\Omega^{(v)}_{\tau} - \hat{\bm{\omega}}^{(v)}_{\tau} \right)  \\
		& + P_{sp}  \delta\left( \Omega^{(v)}_{\tau} - \Bar{\theta} \right).
		\end{split}
		\end{equation}
		
		\textit{Approximation 3:} We assume that if $\hat{\bm{\alpha}}_\tau$, $\hat{\bm{\omega}}_{\tau}$ and $\bm{\pi}_\tau$ are known, then the mean belief parameters $\hat{\bm{\alpha}}_{\tau+1} = \mathbb{E} \{ X_{\tau+1} \mid \bm{\alpha}_t, \mathbf{a}_t \}$ can be computed. Given this assumption, we approximate the conditional probability density over $X_{\tau+1}$ to be degenerate:
		\begin{equation}
		f(X_{\tau+1} \mid \bm{\alpha}_t, \mathbf{a}_t) \approx \delta(X_{\tau+1} - \hat{\bm{\alpha}}_{\tau+1}).
		\end{equation}
		
		It can be noticed that if $X_t = \bm{\alpha}_t$ is observed, then using the aforementioned approximations and the influence diagram, the expected belief parameters $\hat{\bm{\alpha}}_T = \mathbbm{E}_{\psi} \{X_T\mid \bm{\alpha}_{t}, \mathbf{a}_t\}$ can be determined in an iterative manner. Next, we develop centralized algorithms using the framework described so far.
		
		\begin{figure}[h]
			\vspace{-1mm}
			\centering
			\includegraphics[width=0.35\textwidth]{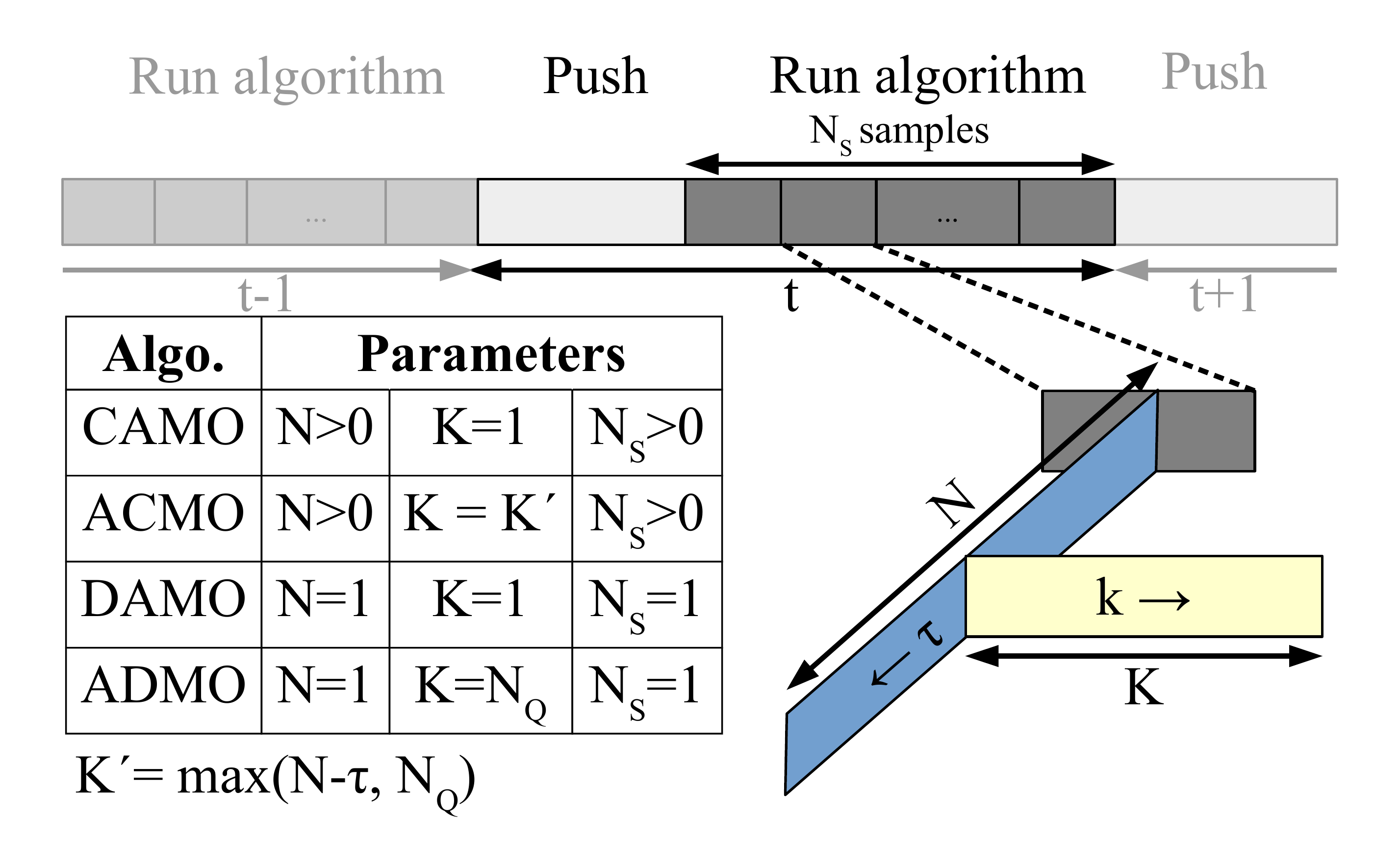}
			\vspace{-2mm}
			\caption{Timeline for different algorithms.}
			\label{fig:timeline}
		\end{figure}
		
		\subsection{CAMO Algorithm}
		\label{sec:camo}
		
		We develop a Centralized Algorithm for Opinion Maximization (CAMO), where we clarify the underlying assumptions mentioned in \ref{sec:framework} by providing some heuristics. First, note that at time $t$ computing the objective function $\mathbbm{1}^T g\left(\mathbbm{E}_\psi\{X_T \mid \bm{\alpha}_t, \mathbf{a}_t \}\right) = \mathbbm{1}^T g(\hat{\bm{\alpha}}_T)$ involves $T-t$ steps of computations. This gives rise to two issues: 1) Large $T-t$ results in the accumulation of error due to approximations at each step. 2) Computational complexity. To address these, we determine the action $\mathbf{a}^*_t$ such that the objective function is maximized for time $t+N$ instead of that at time $T$, where $N < T$ is the look-ahead window size. To achieve this, the state of the network is observed at time $t$, followed by a centralized offline N-step look-ahead procedure. In this regard, we call $t$ as the online time during which the users in the network communicate, the iterations of the algorithm are indexed by $\tau$ (offline time) and $t;\tau$ indicates the composite time. The timelines for different algorithms are depicted in Fig. \ref{fig:timeline}. To be precise, given the state of the network at time $t$, the action $\mathbf{a}_{t;0}$ must be determined such that the mean total opinion at time $t;N$ is maximized. Note that the mean total opinion at $t;N$ is the prediction of the actual future mean total opinion at time $t+N$. Our algorithm consists of the following three stages\footnote{Henceforth, in this section (except Algorithm \ref{algo:nsteplookahead}), $t$ is omitted from $t;\tau$.}:
		\begin{enumerate}
			\item Sampling joint actions $\mathbf{a}_t$.
			\item Computing the objective function $\mathbbm{1}^T g\left(\hat{\bm{\alpha}}_T\right)$ by probabilistic diffusion.
			\item Repeating the previous two steps $N_S$ times, and then choosing sub-optimal joint action $\mathbf{a}^*_t$.
		\end{enumerate}
		
		Next, the aforementioned steps are discussed in detail and the assumptions made in \ref{sec:framework} are addressed.
		\subsubsection{Sampling Joint Actions} \hspace{1mm} \\
		\textit{1.a) Obtaining $\hat{\bm{\omega}}_{\tau}$:}
		Now, we address the assumption associated with Approximation 2. Computing the exact $\hat{\bm{\omega}}_{\tau}$ = $[\hat{\omega}^{(v)}_{\tau}]_{v \in V}$ is tedious since the size of the sample space grows exponentially $\left(|U_\Theta|^{|V|}\right)$ with the network size. Therefore, we assume that every node pushes a message of the class corresponding to its maximum opinion at time $\tau$. Hence, the MAP estimate $\hat{\omega}^{(v)}_{\tau}$ is obtained approximately $\forall v\in V$ and $\forall \tau \in [1,N-1]$ as:
		\begin{equation}
		\hat{\omega}^{(v)}_{\tau} \approx \theta \in \Theta \mid \hat{\alpha}^{(v)}_{\theta, \tau} \geq \hat{\alpha}^{(v)}_{\theta^\prime, \tau}, \Theta \ni \forall \theta^\prime \neq \theta.
		\label{eq:omega*}
		\end{equation}
		However, at time $\tau = 0$, we assume that central controller has the instantaneous knowledge of $\bm{\omega}_{0}$.\\
		\textit{1.b) Obtaining $\bm{\pi}_\tau$:}
		
		According to Approximation 1, the actions of the nodes are independent. Therefore, we can focus on finding mixed strategy for each node, separately. Also, from Remark \ref{remarkBoltz}, we know that mixed strategy can be determined by using individual rewards and Boltzmann distribution. In CAMO algorithm, the individual reward of a node $v$ is the change in opinion of its immediate neighbor $w \in N(v)$ (myopic) caused by pushing message of class $\Tilde{\theta}$ (similar to that in the toy model). Therefore, the mixed strategy of node $v$ is given by:
		\begin{equation}
		\hspace{-2mm}
		\pi_\tau(A^{(v)}_\tau = w) = \begin{cases}
		\Xi(r^{(v)}_{\tau}, w, \mathcal{T}), &\text{ if } \omega^{(v)*}_{\tau} = \Tilde{\theta},
		\\ \frac{1}{|\mathcal{N}(v)|}, &\text{ if } \omega^{(v)*}_{\tau} \neq \Tilde{\theta},
		\end{cases}
		\label{eq:actionprob1}
		\end{equation}
		\noindent where $r^{(v)}_{\tau} = (r^{(v)}_{\tau}(x))_{x \in N(v)}$. Then, the probability over joint actions $\bm{\pi}_\tau$ can be computed using Eq.~\eqref{eq:indep}. As mentioned in Remark \ref{rmk:samp}, the joint action $\mathbf{a}_{0}$ is sampled from $\bm{\pi}_{0}$. Then, to compute the objective function $\mathbbm{1}^T g\left(\hat{\bm{\alpha}}_T\right)$, the central controller performs probabilistic diffusion, which is described as follows.
		
		\begin{algorithm}[h]
			\SetAlgoLined
			{\footnotesize
				Initialize feeds: \\$F^{(v)}_t = (f_i)_{1 \leq i 
					\leq L}$, $\forall v \in V$, where $\mathcal{I}(f_i) \in \Bar{\Theta}$, and $\forall i \in \{1,...,L\}$ \\
				\For{each $t \in \{1,2,..,T\}$}{
					Report $\{\alpha^{(v)}_{\theta, t}\}_{1 \leq \theta \leq \Theta}$ and $\omega^{(v)}_t$ to the central controller, $\forall v \in V$ (For CAMO and ACMO only).\\
					Run learning algorithm: CAMO/DAMO/ADMO/ACMO.\\
					Execute action $\mathbf{a}^*_{t;0}$.\\
					Update belief using Eq.~\eqref{eq:dirichletPredict}.\\
					Update Feed.
				}
				\caption{{\small Wrapper Function.}}
				\label{algo:cen}
			}
			\vspace{3mm}
		\end{algorithm}
		
		\begin{algorithm}[h]
			{\footnotesize
				\SetAlgoLined
				\For{$k \in \{1,2,...,N_S\}$}{
					Compute $\bm{\pi}_{t;0}$ using Eq.~\eqref{eq:pi0}. \\
					Sample $\mathbf{a}_{t;0, n}$ from $\bm{\pi}_{t;0}$ $\left(\mathbf{a}_{t;0, n} \sim \bm{\pi}_{t;0} \right)$.\\
					Initialization: \\ $\hat{\bm{\alpha}}_{t;0} := \bm{\alpha}_{t}$, $\chi^{\text{max}}_{t;N} := 0$, and compute $\hat{\bm{\omega}}_{t; 0}$ using Eq.~\eqref{eq:omega*}.\\
					\For{each $\tau \in \{1,2,...,N\} $}	    
					{
						Compute $\hat{\bm{\omega}}_{t;\tau}$ using Eq.~\eqref{eq:omega*}.\\
						Compute $\hat{\bm{\alpha}}_{t;\tau}$ using Eq.~\eqref{eq:alphacap}.\\
						\textbf{CAMO:}\\
						\hspace{2mm} Compute $\bm{\pi}_{t;\tau}$ using Eq.~\eqref{eq:actionprob1}.\\
						\textbf{ACMO:}\\
						\hspace{2mm} Run ADMO to get $\bm{\pi}_{t;\tau} = \left(\Xi(Q^{(u)}_{t;\tau;K}, v, \mathcal{T})\right)_{\substack{u \in V \setminus V_r \\ v \in \mathcal{N}(u)}}$.			    
					}
					Compute $\chi_{t,N} = \mathbbm{1}^Tg(\hat{\bm{\alpha}}_{t;N})$.\\
					\If{$\chi_{t;N} > \chi^{\text{max}}_{t;N}$}
					{
						$\mathbf{a}^*_{t;0} := \mathbf{a}_{t;0, n}$. \\ 
						$\chi^{\text{max}}_{t;N} := \chi_{t;N}$.
					}
				}
				\caption{{\small CAMO and ACMO Algorithms.}}
				\label{algo:nsteplookahead}
			}
		\end{algorithm}
		
		\subsubsection{Probabilistic Diffusion} 
		
		Given $\bm{\alpha}_0$ and $\mathbf{a}_0$, computing $\hat{\bm{\alpha}}_N$ iteratively is termed as \textit{probabilistic diffusion}, since belief parameters (hence opinions) evolve probabilistically in the network through information spreading. To complete the steps involved in probabilistic diffusion, we address the assumption associated with Approximation 3, by deriving the expression for $\hat{\bm{\alpha}}_{\tau+1}$ as follows. \\
		\textit{Computing $\hat{\bm{\alpha}}_{\tau+1}$:}
		\begin{equation}
		\begin{split}
		\hat{\bm{\alpha}}_{\tau+1} & = \mathbb{E}\{X_{\tau+1} \mid \bm{\alpha}_0, \mathbf{a}_0 \} = \mathbb{E}\{\beta X_{\tau} + \Delta X_{\tau} \mid \bm{\alpha}_0, \mathbf{a}_0 \}\\
		& = \bm{\beta} \circ \hat{\bm{\alpha}}_{\tau} + [\mathbb{E}\{\Delta X^{(v)}_{\theta, \tau} \mid \bm{\alpha}_0, \mathbf{a}_0 \}]_{\substack{v \in V \\ \theta \in \Theta}},
		\end{split}	
		\label{eq:alphacap}
		\end{equation}	
		where\footnote{$\circ$ denotes the Hadamard product.} $\bm{\beta} = [\beta^{(v)}]_{v \in V}$, and
		\begin{equation}
		\begin{split}
		\mathbbm{E}&\left\{\Delta X^{(v)}_{\theta, \tau} \mid \bm{\alpha}_0, \mathbf{a}_0\right\} \\
		& = \zeta^{(v)} \hspace{-2mm} \sum\limits_{u \in N(v)} \mu^{(u)}_{\theta, \tau} \pi_{\tau}(A^{(u)}_\tau = v) (1-P_{sp}) \delta(\theta - \hat{{\omega}}^{(v)}_\tau).
		\end{split}
		\label{eq:delblf}
		\end{equation}	
		The derivation of Eq.~\eqref{eq:delblf} is given in Appendix \ref{app:delblfderivn}. To determine $\bm{\hat{\alpha}}_{\tau+1}$ for $\tau > 0$, $\bm{\pi}_\tau$ is computed as given in Eq.~\eqref{eq:actionprob1}. However, to compute $\hat{\bm{\alpha}}_1$ the conditional probability $\bm{\pi}_{0}$ is modified as:
		\begin{equation}
		\pi_0(A^{(u)}_{0} = v) = \begin{cases}
		\delta(v - a^{(u)}_{0}), &\text{ if } \theta = \Tilde{\theta},
		\\
		\frac{1}{|\mathcal{N}(u)|}, & \text{ if } \theta \neq \Tilde{\theta}.
		\end{cases}
		\label{eq:pi0}
		\end{equation}
		In other words, a node $u$ which has chosen message of class $\Tilde{\theta}$ from its feed, pushes message to node $a^{(u)}_{0}$, whereas a node that has chosen a message corresponding to a random source selects one of their neighbors u.a.r.
		
		\subsubsection{Choosing Sub-optimal Action $\mathbf{a}_{0}^*$}
		Let $[\mathbf{a}_{0, n}]_{1 \leq n \leq N_{S}}$ be the actions sampled from $\bm{\pi}_{0}$ and $[\hat{\bm{\alpha}}_{N,n}]_{1 \leq k \leq N_{S}}$ be the belief parameters at time $\tau = N$. Then the sub-optimal action $\mathbf{a}_{0}^*$ is chosen as: $\mathbf{a}^*_{0} = \underset{\mathbf{a}_{0, n}}{\text{argmax}} \text{ } \mathbbm{1}^T g(\hat{\bm{\alpha}}_{N, n})$.
		\subsection{ACMO Algorithm}
		Augmented Centralized algorithm for Opinion Maximization (ACMO) is an improved variant of the CAMO algorithm which is made to piggy-back on ADMO (described in \ref{sec:admo}). The main limitation of CAMO algorithm that the individual rewards are computed in a myopic manner. To alleviate this problem, at each offline time $\tau$, Q-learning is used to look-ahead in time to compute the individual rewards, and hence obtain better mixed strategies. More precisely, $\Xi(r^{(v)}_{\tau}, w, \mathcal{T})$ in Eq.~\eqref{eq:actionprob1} is replaced by $\Xi(Q^{(v)}_{\tau;K}, w, \mathcal{T})$, where $K = \text{max}(N-\tau, N_Q)$, and obtaining $Q^{(v)}_{\tau;K}$ is described in \ref{sec:admo}. Note that the composite time consists of an additional offline time $k$ (depicted in Fig. \ref{fig:timeline}) to capture Q-learning iterations. Algorithm \ref{algo:cen} is the wrapper function, which is a general pseudo-code common for all algorithms. The CAMO and ACMO algorithms are given in Algorithm \ref{algo:nsteplookahead}.
		
		\section{Decentralized Algorithms}
		\label{sec:decentraliz}
		\noindent In this section, two different variants of decentralized algorithms are presented: \begin{inparaenum} \item Decentralized Algorithm for Opinion Maximization (DAMO), and \item Augmented Decentralized algorithm for Opinion Maximization (ADMO). \end{inparaenum}
		\subsection{DAMO Algorithm}
		DAMO is a special case of CAMO algorithm obtained by using $N_S = 1$ and the window size $N = 1$. Note that in centralized algorithms, a central controller is required for probabilistic diffusion and to store joint action-future reward pairs obtained by repeated sampling of joint actions. Setting $N_S = 1$ implies that the every node samples the action independently from its mixed strategy only once, and $N=1$ implies that there is no probabilistic diffusion. This makes the algorithm decentralized. Hence, the action taken at time $t$ is simply, $\mathbf{a}^*_t \sim \bm{\pi}_t$. The algorithm admits a simple two-step procedure given in Algorithm \ref{algo:damo}.
		\begin{algorithm}
			{\footnotesize
				\SetAlgoLined
				Compute $\Xi(r_t^{(u)}, v, \mathcal{T})$, $\forall u \in V \setminus V_r$. \\
				\textbf{DAMO:} Sample $a_t^{(u)} \sim \left(\Xi(r_t^{(u)}, v, \mathcal{T})\right)_{v \in \mathcal{N}(u) \setminus u}$, $\forall u \in V \setminus V_r$. \\
				\caption{{\small DAMO Algorithm.}}
				\label{algo:damo}
			}
		\end{algorithm}
		
		Next, a brief background on Q-learning is provided, since it manifests itself in the ADMO algorithm discussed subsequently.
		\subsection{Background on Q-Learning}
		\textit{Q-learning} is a model-free reinforcement learning algorithm. For any Markov Decision Process, Q-learning can be used to find the optimal policy, which is obtained by learning the so-called action-value function $Q(s, a)$, where $a$ is the action taken when the system is in state $s$. In particular, $Q(.,.)$ is the expected discounted reward of taking action $a$ in state $s$ and continuing optimally thereafter \cite{watkins1992q}.
		When an agent is in state $s_t$, the probability of taking action $a_t$ is given by: ${P}(s_t, a_t) = \Xi(Q(s_t), a_t, \mathcal{T})$, where $\mathcal{T}$ is the temperature parameter and $Q(s_t) = (Q(s_t, a^\prime))_{a^\prime}$. If $r_t(s_t, a_t)$ is the instantaneous reward obtained at time $t$ by taking action $a_t$ when the system is in state $s_t$, then the Bellman-equation to update action-values is given by:
		\begin{equation}
		\hspace{-0.5mm} 
		Q_{t+1}(s_t, a_t) \hspace{-0.3mm} := \hspace{-0.3mm} \Bar{\lambda} Q_t(s_t, a_t) + \lambda [r_t(s_t, a_t) + \gamma \underset{a^\prime}{\max}Q_t(s_t, a^\prime)],\hspace{-0.5mm}
		\label{eq:qlrnstate}
		\end{equation}
		where $\Bar{ \lambda} = 1-\lambda$ and $\gamma \in [0,1]$ is the discount factor.
		\label{sec:qlrn}
				
		\subsection{ADMO Algorithm}
		\label{sec:admo}		
		The basic idea of the ADMO algorithm can be illustrated using an example: Consider a node $v$ that has two neighbors $w_1$ and $w_2$. The node $w_1$ can be persuaded easily, but has a few stubborn neighbors. On the other hand, node $w_2$ is hard to persuade, but has a large number of persuadable neighbors. Given such a scenario, in the DAMO algorithm node $v$ pushes the message to $w_1$ myopically. However, despite the immediate reward (change in opinion) being lower, it would be wiser to persuade $w_2$ because it is more influential, and hence would yield higher reward after a few time steps.  In the ADMO algorithm, each node selfishly looks ahead in time by exploring beyond neighbors over multiple hops for better rewards, based on which better strategies are determined. 
		
		We develop the ADMO algorithm based on the idea presented for the simplified model in Appendix \ref{ap:prop1}, where a single message circulates in the network and the environment is static. However, in contrast to the simplified model, we observe three differences about opinion dynamics in the actual model described in \ref{sec:commodel}.
		\begin{inparaenum}
			\item There are multiple messages circulating in the network. Hence, every node experiences a dynamic environment due to the change in opinions caused by transmissions of other nodes.
			\item Initially a few nodes would be transmitting messages of class $\Theta$. Hence, the opinions of many nodes remains unchanged. Therefore, the environment can be considered to be slowly varying during initial time steps.
			\item On the other hand, as time progresses, more number of nodes would be transmitting messages of class class $\Theta$ rendering the environment more dynamic.
		\end{inparaenum}
		Considering the aforementioned observations, we introduce a time varying discount factor $\gamma_t = \gamma^\prime \gamma^{\prime \prime^t}$, where $\gamma^{\prime}, \gamma^{\prime \prime} \in [0,1]$, to weigh down the future rewards. Note that the discount factor decays with time $t$ to account for the environment becoming increasingly dynamic with time. In this algorithm, each node uses the sum of discounted future rewards (s.o.d.f.r.) as individual rewards to determine its mixed strategy independently. The state of the network is observed at time $t$ and s.o.d.f.r. is computed iteratively. We denote the iteration number (learning time) as $k$ and the composite time as $t;k$. $\forall v\in V|\mathcal{I}(m^{(v)}_{t;k} = \Tilde{\theta})$ the s.o.d.f.r. is maximized over the residual time $T-t$, which is given as (omitting the $t$ in the composite time $t;k$ without loss of generality):
		\begin{equation}
		\underset{\left(a^{(u_k)}_k\right)_{0\leq k\leq T-1-t}}{\text{max}} \sum\limits_{k=0}^{T-1-t}
		\gamma^{k}_{t} r^{(u_k)}_k(a^{(u_k)}_k),
		\label{eq:decenrwddisc}
		\end{equation}
		where $u_0 = v$. We assume that every node in the network independently attempts to maximize its sum of future discounted rewards over a finite time horizon $T-1-t$. In this respect, the sum of future discounted rewards of node $v$ when $a^{(v)}_0 = x$ is given by:
		\begin{equation}
		Q^{(v)}_{T-1-t}(x) = r^{(v)}_0(x) + \hspace{-2mm}
		\underset{\left(a^{(u_k)}_k\right)_{1\leq k\leq T-1-t}}{\text{max}} \hspace{-2mm} \sum\limits_{k=1}^{T-1-t}
		\gamma^{k}_{t} r^{u_k}_k(a^{(u_k)}_k),
		\label{eq:decenrwddiscQ1}	    	    
		\end{equation}	
		where $r^{(u)}_0(v)$ is the reward obtained upon pushing message to node $v$ at time $k=0$. We generalize Eq.~\eqref{eq:decenrwddiscQ1} for $0 \leq l \leq T-1-t$ as:
		\begin{equation}
		Q^{(v)}_{l+1}(x) =  r^{(v)}_{0}(x) + \gamma_t Q_{{l}_{max}}^{(x)}(v),
		\label{eq:qlrn}
		\end{equation}
		where $Q_{{l}_{max}}^{(x)}(v) = \underset{w \in \mathcal{N}(x) \setminus v}{\max} Q_{l}^{(x)}(w)$. Node $v$ is excluded from the action set of node $x$ to avoid back-and-forth influence between a pair of nodes. By comparing Eq.~\eqref{eq:qlrn} with Eq.~\eqref{eq:qlrnstate}, it can be observed that each node employs stateless Q-learning. In this algorithm, each node determines the individual rewards (s.o.f.d.r.) by exchanging the action-values between the nodes. For instance, in Eq.~\eqref{eq:qlrn} node $x$ shares $Q^{(x)}_{l_{max}}(v)$ with node $v$, which consequently updates its action-value $Q_{l+1}^{(v)}(x)$. Also, to alleviate the computational complexity, the action-values are updated only for $N_Q$ time steps. The ADMO algorithm is given in Algorithm \ref{algo:admo}.
		\begin{algorithm}
			{\footnotesize
				\SetAlgoLined
				\textbf{Initialization}:
				$Q^{(u)}(v) := 0$, $\forall u \in V \setminus V_r$ \text{, and } $\forall v \in \mathcal{N}(u)$. \\
				$k := 0$.\\
				\While{$k < N_Q$} {
					\vspace{-5mm}
					\begin{multline}
					\hspace{-1cm}
					Q^{(u)}_{t;k+1}(v) := r^{(u)}_{t; 0}(v) + \gamma^{k}_{t} Q_{t;k_{max}}^{(v)}(u), \forall u \in  V \setminus V_r \text{, and } \forall v \in \mathcal{N}(u) \setminus u.  \notag
					\end{multline}		    
				}
				Compute $\Xi(Q^{(u)}_{t;N_Q}, v, \mathcal{T})$, $\forall u \in V \setminus V_r$. \\
				Sample $a_t^{(u)} \sim \left(\Xi(Q^{(u)}_{t;N_Q}, v, \mathcal{T})\right)_{v \in \mathcal{N}(u)}$, $\forall u \in V \setminus V_r$.
				\caption{{\small ADMO Algorithm.}}
				\label{algo:admo}
			}
		\end{algorithm}
		\section{Complexity Analysis}
		\label{sec:complexity}    
		\noindent The required knowledge as well as the space and time complexities of different variants of centralized and decentralized algorithms are listed in Table \ref{tbl:complexity}. In the decentralized algorithms, each node must know the opinions, the overall strength of the opinions and the social learning abilities of only the neighboring nodes. For a given information spreading process (equivalently the source node), the opinion of the neighbors corresponding to only its class must be known. On the other hand, the centralized algorithm requires that the opinions pertaining to all the message classes be known. Moreover, centralized algorithm requires the knowledge of the topology of the graph $G$, and the locations of all the source nodes $V_S$. Therefore, the centralized algorithms bear a significant overhead compared to the decentralized variants.
		{\renewcommand{\arraystretch}{1.4}
			\begin{table}
				\begin{center}
					\resizebox{\columnwidth}{!}{
						\begin{tabular}{|l|l|l|l|l|}
							\hline
							& \textbf{Algorithm} & \textbf{Space}  & \textbf{Time}  & \textbf{Required}  \\
							& &  \textbf{complexity} &  \textbf{complexity} &  \textbf{knowledge} $(\forall v \in V)$ \\
							\hline
							\parbox[t]{1mm}{\multirow{2}{*}{\rotatebox[origin=c]{90}{{\footnotesize Cen.}}}} &
							\hspace{0.1cm} 1. CAMO & $O(|E|)$ & $O(|E| N_S N)$  &  $[\mu^{(v)}_\theta]_{\theta \in \Theta}$, $\rho^{(v)}$, $P_{sp}$ \\
							\cline{2-4}
							& \hspace{0.1cm} 2. ACMO  & $O(|E|)$  & $O(|E| N_Q N_S N)$ & $\beta^{(v)}$, $\zeta^{(v)}$, $G$, $V_S$. \\
							\hline
							\parbox[t]{1mm}{\multirow{3}{*}{\rotatebox[origin=c]{90}{{\footnotesize \hspace{3mm} Decen.}}}} &
							\hspace{0.1cm} 3. DAMO & $O(d_{max})$ & $O(d_{max})$ & $\rho^{(w)}$, $\mu^{(w)}_{\Tilde{\theta}}$, $\beta^{(w)}$, $\zeta^{(w)}$, \\
							\cline{2-4}
							& \hspace{0.1cm} 4. ADMO & $O(d_{max})$ & $O(d_{max} N_Q)$ & $\forall w \in \mathcal{N}(v)$.
							\\
							\hline
						\end{tabular}
						}
					\end{center}
				\caption{Complexity and required knowledge of the algorithms.}
				\label{tbl:complexity}
				\vspace{-5mm}
			\end{table}
		}
		In the centralized algorithms, the space complexity is dominated by the storage of action-values of all the nodes, where each regular node $v$ has $|N(v)|$ action-values resulting in $2|E|$ action-values for the entire network. This leads to the space complexity of $O\left(|E|\right)$. On the other hand, the space complexity of the decentralized algorithm is predominantly due to the individual action-values, which results in the space complexity of $O\left(d_{max}\right)$, where $d_{max}$ is the maximum degree in the network.
		
		The time complexity of the ACMO algorithm is dominated by the probabilistic diffusion and repeated Q-learning. Both probabilistic diffusion and one-step Q-learning involve about $|E|$ operations for every offline time $\tau$. Considering N-step look ahead (window of size $N$), $N_Q$ repetitions of Q-learning and $N_{S}$ samples, the number of operations per unit time ($t$) of the centralized algorithm is $O(|E|N_Q N_{S}N)$. Since the CAMO algorithm is similar to the ACMO algorithm without Q-learning, its time complexity is $O(|E|N_{S}N)$. The decentralized algorithms have much lower time complexity. In ADMO algorithm, each node $v \in V$ performs repeated Q-learning independently, which involves $|\mathcal{N}(v)| N_Q$ operations per unit time resulting in the worst-case time complexity of $O(d_{max} N_Q)$. Since, the DAMO algorithm does not involve Q-learning, its time complexity is $O(d_{max})$.
		\section{Simulation Results}
		
		\begin{figure*}[t]
			\minipage{5.5cm}
			\subcaptionbox{Final total opinion of the population versus the centrality ($C_{cl}$) of the smart source for different algorithms.\label{fig:boxplot}}[0.97\linewidth][c]{\includegraphics[width=.97\linewidth, height=0.56\linewidth]{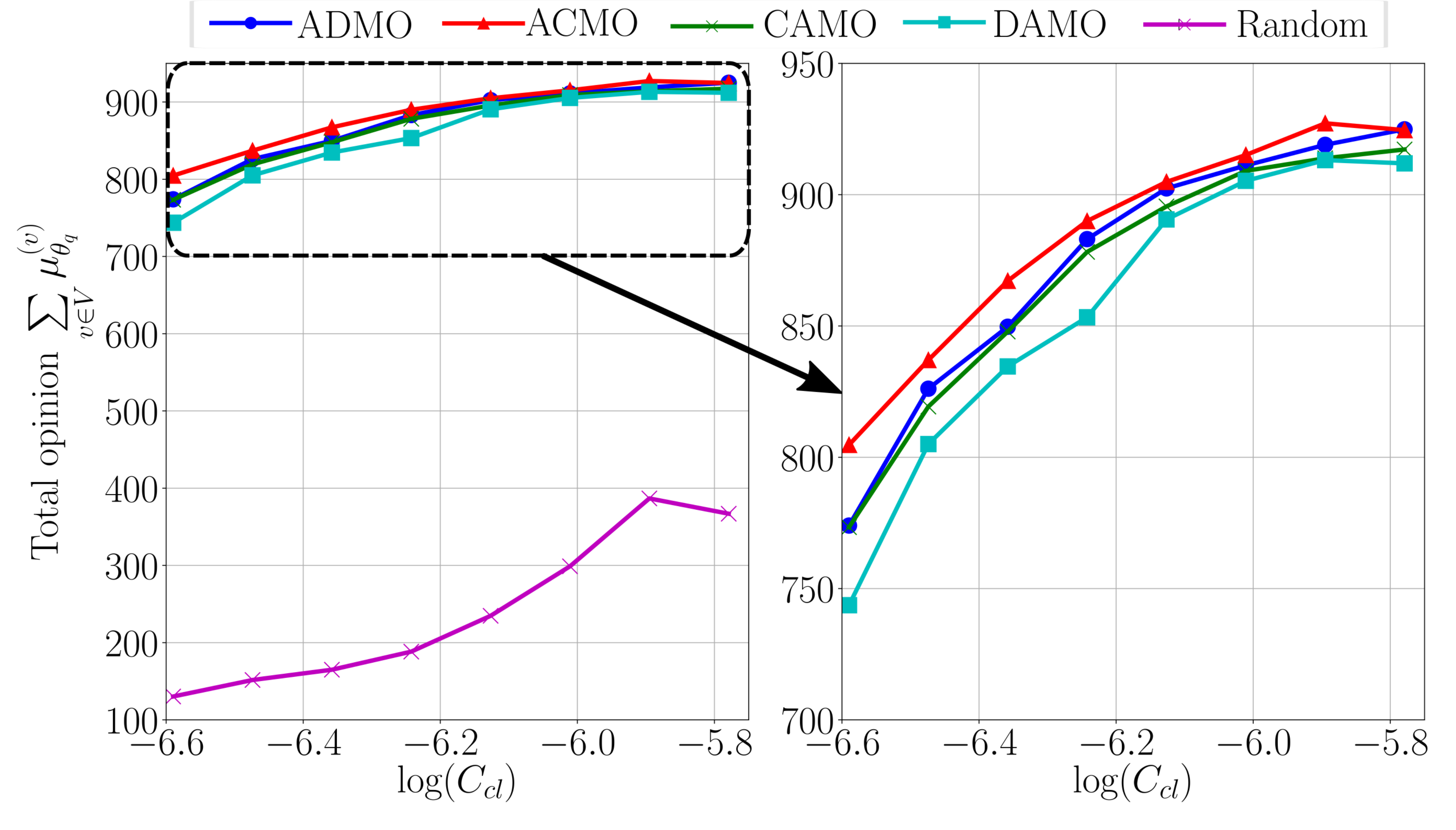}}
			\subcaptionbox{Evolution of total opinion with time: for different algorithms (left), and for $3$ different classes when ADMO is used (right).\label{fig:sumopinall}}[.97\linewidth][c]{\includegraphics[width=.97\linewidth, height=0.56\linewidth]{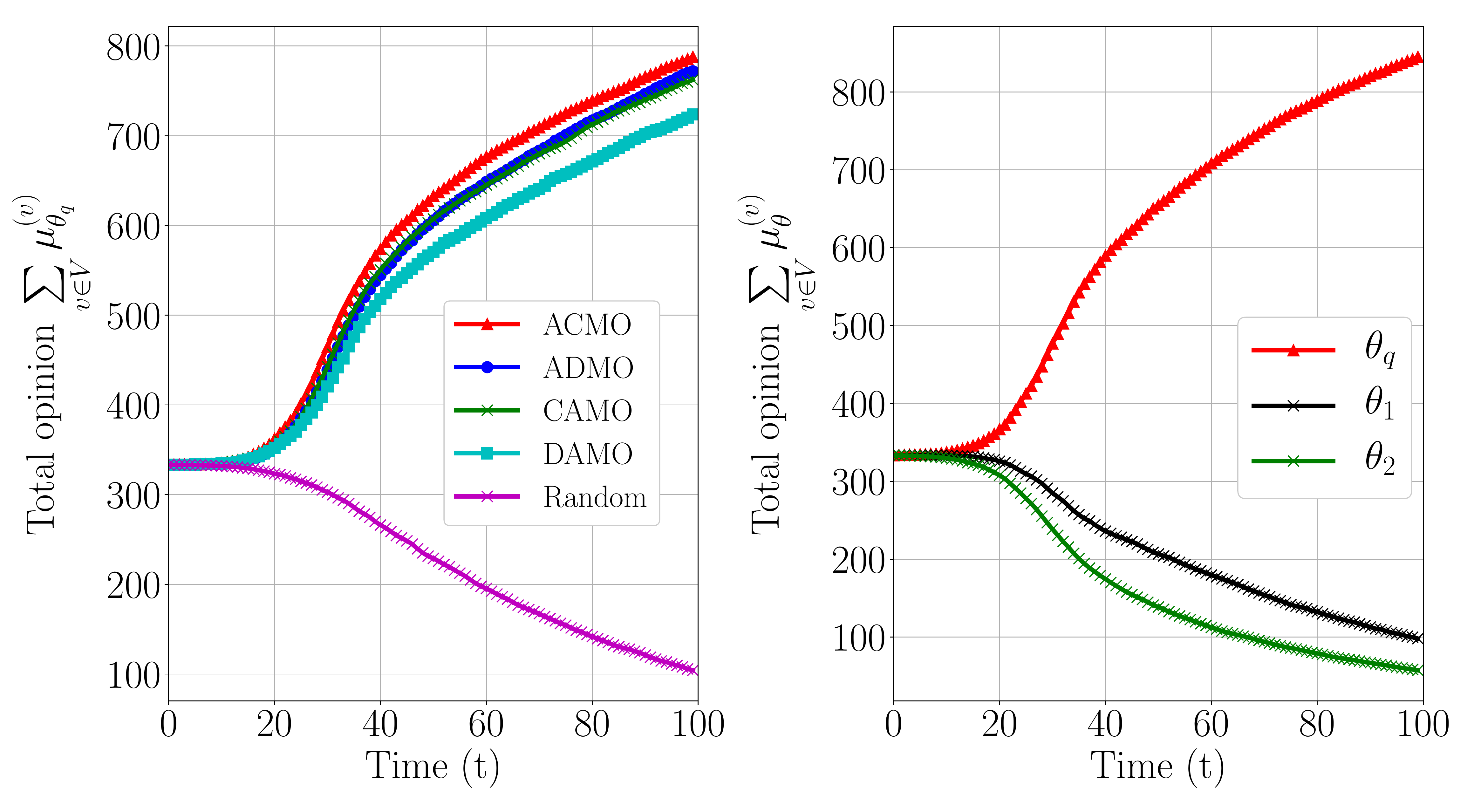}}
			\caption{Effect of centrality and evolution of total opinion for different algorithms in PA graph of $10^3$ nodes.}
			\endminipage
			\quad
			\minipage{5.5cm}        
			\subcaptionbox{Evolution of total opinion with time (left), and total opinion at time $t=100$ (right) for different algorithms in PA graph of $10^4$ nodes.\label{fig:10k}}[.97\linewidth][c]{\includegraphics[width=.97\linewidth, height=0.56\linewidth]{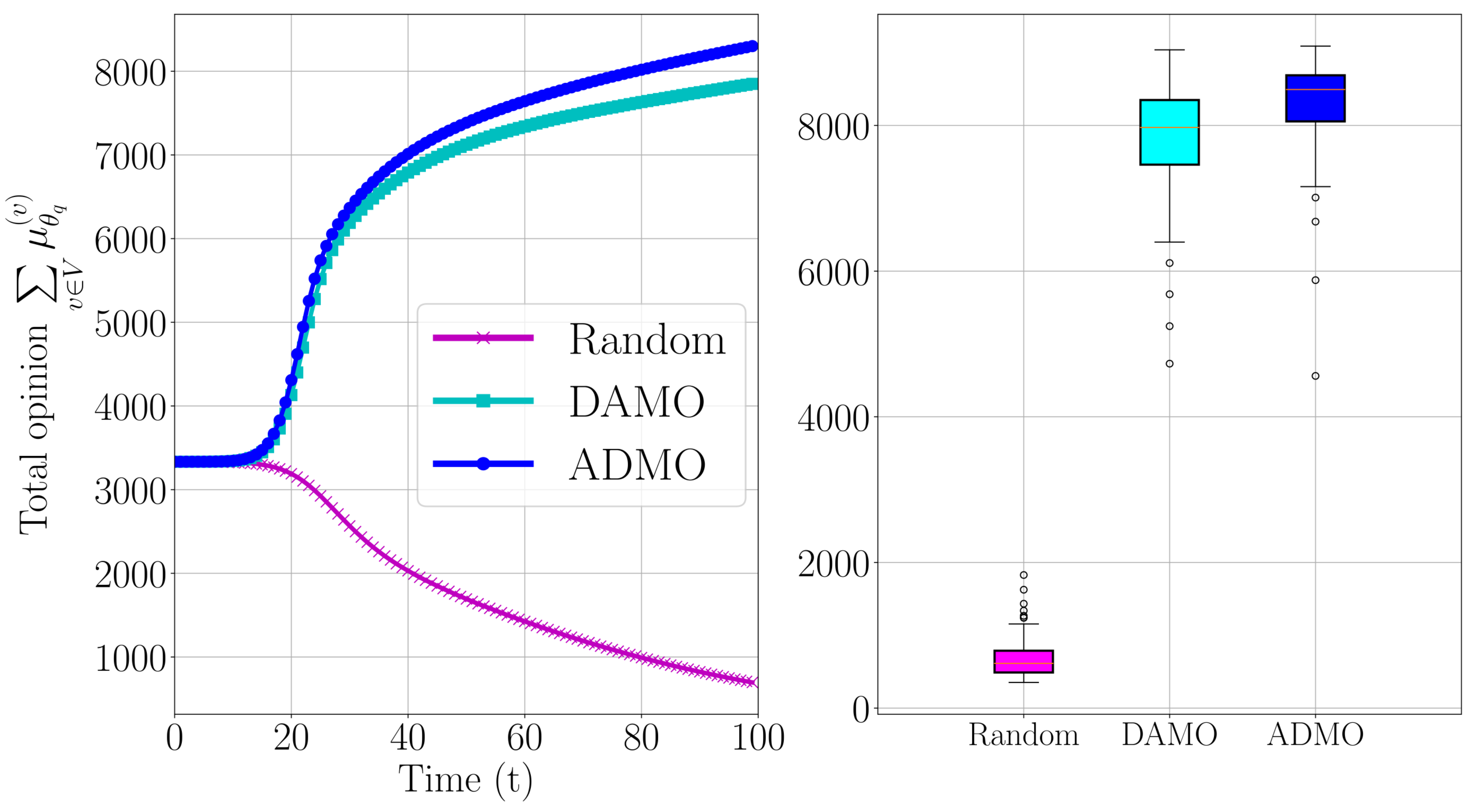}}
			\subcaptionbox{Evolution of total opinion with time (left), and total opinion at time $t=100$ (right) for different algorithms in Facebook ego-network.\label{fig:fbego}}[.97\linewidth][c]{\includegraphics[width=.97\linewidth,height=0.56\linewidth]{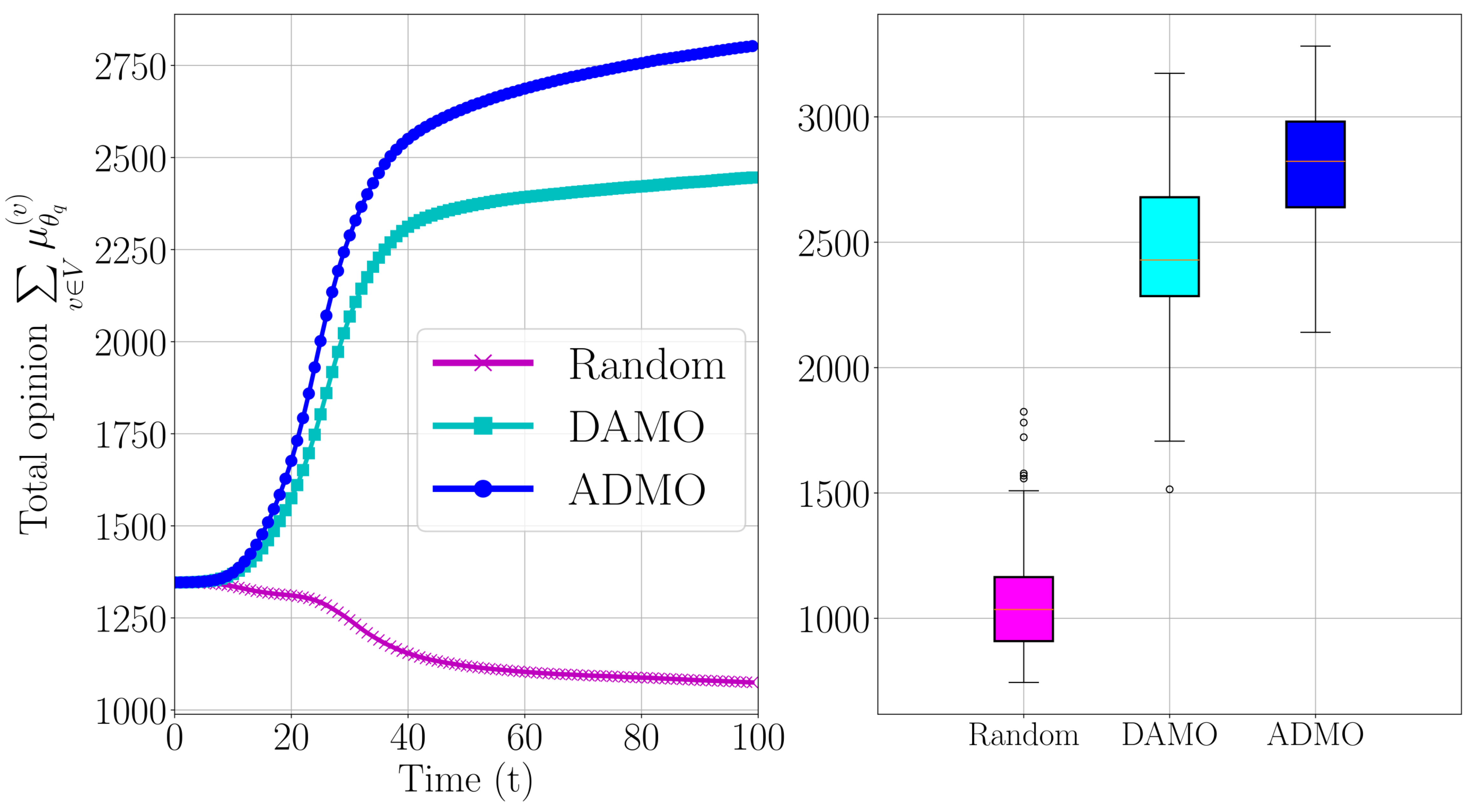}} 
			\caption{Evolution of total opinion and final opinion for different algorithms in PA graph and Facebook ego-network.}
			\endminipage
			\quad
			\minipage{5.5cm}
			\subcaptionbox{Visualization of opinions in PA graph of $10^3$ nodes (top) and Facebook ego-network (bottom).
				Blue - affinity towards $\Tilde{\theta}$, and green and red - affinity towards $\Theta \setminus \Tilde{\theta}$.\label{fig:viz}}[1.15\linewidth][c]{
				\includegraphics[width=.8\linewidth, height=0.56\linewidth]{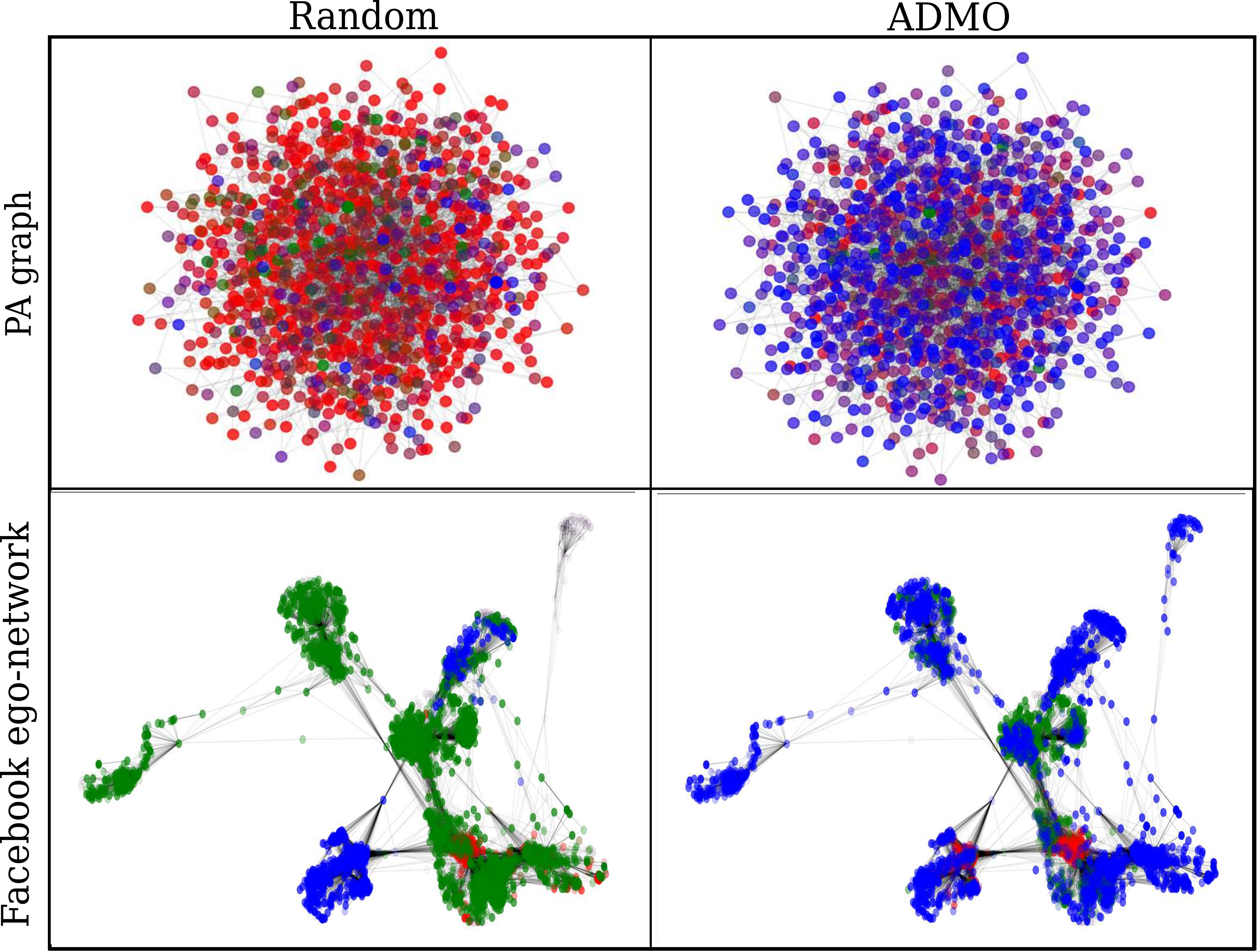}
			}
			\subcaptionbox{Evolution of beliefs at time $t=20, 30, 50$ and $100$ (clock-wise), respectively, in PA graph of $10^3$ nodes.
				\\ \hspace{1mm}
				\label{fig:blfevol}}[1.15\linewidth][c]{
				\includegraphics[width=.6\linewidth, height=0.56\linewidth]{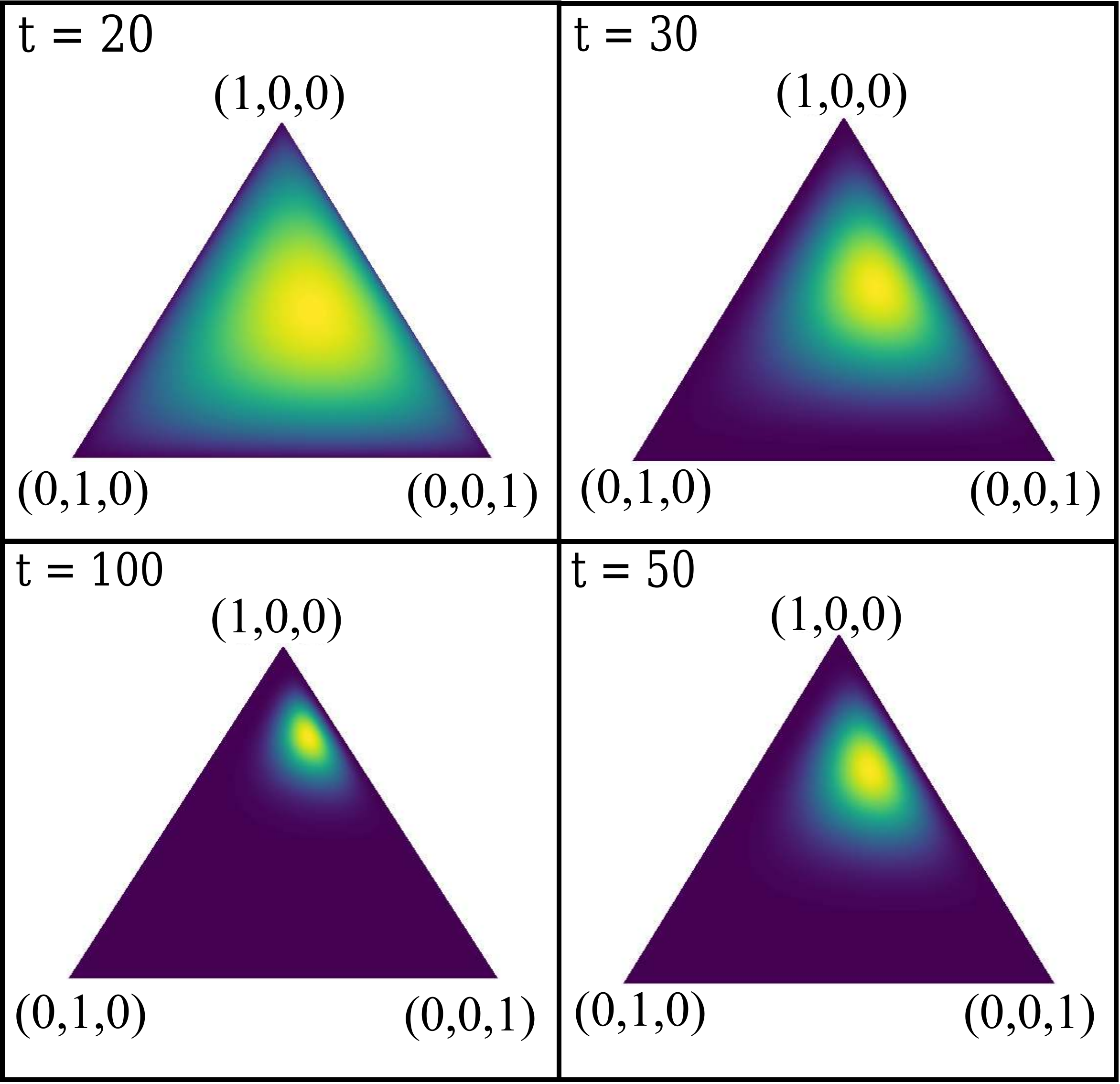}}
			\captionsetup{width=1\linewidth}
			\caption{Visualization of the evolution of opinions in graphs and the temporal evolution of beliefs.}
			\endminipage
			\vspace{-2mm}
		\end{figure*}
		
		\label{sec:simres}
		\noindent We present the simulation results for: 1) Barabasi-Albert preferential-attachment (PA) graph \cite{barabasi1999mean}, and 2) Facebook ego-network \cite{snapnets},\cite{leskovec2012learning}. We consider PA graphs of $10^3$ nodes and $10^4$ nodes with the preferential attachment parameter $m = 3$. Facebook ego-network consists of $4039$ nodes with a high clustering coefficient ($0.6055$) and small diameter ($8$). There are 3 sources in the network, one of which employs the smart information spreading, while the rest spread information at random. The discount factors $\gamma^\prime$ and $\gamma^{\prime \prime}$ are set to $0.95$ and $0.97$, respectively.
		The temperature $\mathcal{T}$ of Boltzmann exploration is set to $0.015$ for $10^3$ nodes and $0.03$ for $10^4$ nodes. The parameters pertaining to the communication model $P_{sp}$, $L$ and $R_m$ are set to $0.1$, $20$, and $2$, respectively. Finally, the remembering factor $\beta$ and the belief update parameter $\zeta$ are uniformly distributed for each node in [0.9, 1] and [0, 2], respectively. The parameters $N_Q$ and $N$ are both set to: 4 for PA graph with $10^3$ nodes, and 5 for PA graph with $10^4$ nodes and Facebook ego-network. For the centralized algorithms, the number of samples $N_S = 20$. All the simulation results are averaged over $100$ iterations. In our simulations, we evaluate the performances of the following: \begin{inparaenum} \item Proposed algorithms: In this case, messages of node $\Tilde{v}$ are spread in the network using one of the proposed algorithms, while the messages of the nodes in $V_r$ are spread randomly. \item Random baseline: Messages of all the sources $V_S$ are pushed by the regular nodes in the network uniformly at random.
		\end{inparaenum}
		Moreover, to mimic the real-world scenario, in the simulations we consider that a node's opinion is unaltered upon reception of duplicate messages.
		\subsection{Final Opinion and Centralities}
		We define the \textit{final opinion} of a node to be its opinion at time $T = 100$. Fig. \ref{fig:boxplot} shows the plot of average final total opinion of the population versus the centrality of the smart source $\Tilde{v}$ in the considered PA graph with $10^3$ nodes. The centralities of random sources are $3.3 \times 10^{-3}$ (hub) and $1.94 \times 10^{-3}$ (intermediate). We have chosen current-flow closeness centrality \cite{brandes2005centrality} since it exhibits the highest correlation with the final total opinion. The Pearson correlation coefficient (PCC) for different choices of centralities is shown in Table \ref{tbl:pearson}. It can be observed that the average opinion of the nodes with smart information spreading is significantly greater than its random counterpart. We can also observe from the figure that even though a node is unfavourably located (away from hubs in PA graph), the opinion maximization can be achieved through smart information spreading using our proposed algorithms.
		\begin{table}
			\begin{center}
				\resizebox{\columnwidth}{!}{
					\begin{tabular}{|c|c|c|c|c|c|}
						\hline
						\hspace{-2mm} \textbf{Centrality} \hspace{-2mm} & \hspace{-2mm} \parbox{1.5cm}{Current-flow \\ closeness} \hspace{-2mm} & \parbox{1.5cm}{Current-flow \\ betweenness} & \hspace{-2mm} Betweenness \hspace{-2mm} & \hspace{-2mm} Closeness \hspace{-2mm} & \hspace{-2mm} Degree \hspace{-2mm} \\
						& & & & & \\
						\hline
						\textbf{PCC} & 0.77 & 0.66 & 0.60 & 0.45 & 0.64 \\
						\hline
					\end{tabular}
				}
				\end{center}
			\caption{PCC for different choices of centralities.}
			\label{tbl:pearson} 
			\vspace{-5mm}
		\end{table}
		
		\subsection{Opinion Evolution with Time}
		Fig. \ref{fig:sumopinall} depicts the evolution of the total opinion of the population with time for different variants of the proposed centralized and decentralized algorithms. The centrality of the smart source $\Tilde{v}$ is $1.5 \times 10^{-3}$. The order of performance can be observed as: ACMO $>$ ADMO,CAMO $>$ DAMO. It can be observed that using Q-learning based approach improves the performance of the algorithms. Since, the beliefs are slowly varying with respect to time, Q-learning can be used to estimate the future reward up to $N_Q$ time steps and actions can be chosen accordingly. Based on the application specific requirements, the decentralized algorithms can be used effectively due to their lower computational complexity.
		In larger networks ($>$ $10^3$ nodes), owing to the higher complexity of the centralized algorithms, the performance of only the ADMO and DAMO algorithms are depicted. Fig. \ref{fig:10k} shows the evolution of total opinion with time corresponding to all the sources (smart and random) in a PA graph with $10^4$ nodes. It can be observed that the smart sources employing learning-based active information spreading process can polarize the opinion of the population, while the random sources influence only a small fraction of nodes. In Fig. \ref{fig:fbego}, the evolution of total opinion and the final total opinion is depicted for the proposed algorithms in the Facebook ego-network with 4039 nodes along with the random baseline. The performance trend is similar to that of PA graphs as described earlier. Moreover, we can observe that the performance gap between DAMO and ADMO algorithms in the Facebook ego-network is larger than that in PA graph owing to the community structure of the Facebook ego-network, which is utilized by ADMO by penetrating outside the community for better rewards. Considerable improvement can be observed in the performance of the proposed algorithms over random spreading. Also, the average final total opinion of ADMO algorithm about 25 percent greater than that of DAMO algorithm.
		
		\subsection{Visualizing Evolution of Beliefs}
		The influence maximization can be visualized in Fig. \ref{fig:viz}, where the goal is to improve the influence of the blue colored smart contagion. The contagions colored green and red are random spreading processes. We have considered a PA graph of $10^3$ nodes and the Facebook ego-network for illustration. Left-half of the figure depicts the influence regions when all the sources employ random spreading process, and the right-half of the figure depicts the opinions when spreading process indicated in blue employs ADMO algorithm. It can be visually observed that by using ADMO algorithm, the influence region has expanded beyond the community (to which the smart source belongs) in the Facebook ego-network, and the fraction of influenced nodes have increased in the PA graph. Fig. \ref{fig:blfevol} shows the evolution of beliefs of the population with time when the smart node uses ADMO algorithm. The triangles shown in the figure depict the support of opinions, and the probability density function over the support is the average belief. The average belief becomes stronger (low variance) and more inclined towards the opinion of the smart node i.e., $(1,0,0)$. The average belief parameters are obtained as: $\alpha^{m}_{j, T} = (\sum_{v \in V} \alpha^{(v)}_{j,T} )/|V|$, $\forall j \in \Theta$.
		
		\section{Conclusion}
		\label{sec:conclus}
		\noindent In this paper, opinion maximization in social networks is formulated and studied from the angle of efficient information spreading using gossip mechanism. We considered a scenario where a smart source employs efficient spreading process against multiple random adversarial sources. The social interactions and opinion dynamics in the network are modeled as a dynamic Bayesian network, using which we formulated the opinion maximization as a sequential decision problem. Due to its intractability, a series of approximations are proposed to develop centralized algorithms. Then, to address the issue of scalability, we proposed online decentralized algorithms with lower computational complexities. The proposed algorithms use learning-based techniques, which facilitate active information spreading. Simulation results are presented for PA graphs and an instance of Facebook graph. In our simulations, we make two important observations: First, the proposed algorithms outperform the baseline (random spreading) by a large margin. Second, even though the source is unfavorably located, using the proposed algorithms it can achieve better performance compared to random information spreading even though its location is favorable.
		
		
		
		\appendices
		
		\section{Analyzing the Opinion Maximization Problem}
		\label{ap:prop1}
		\noindent Obtaining a closed-form expression to maximize the opinion of the population as defined in Definition \ref{def:opinmax} is non-trivial because of the dependency on topology of the graph, centralities of the source nodes, and the non-stationarity of the environment introduced by belief updates. To this end, we obtain a closed-form expression for a special case of the model described in \ref{sec:commodel}, which is defined as follows:
		\begin{definition}
			\textit{Simplified communication model} is obtained by imposing the following constraints $\forall v \in V$ and $\forall t \in \{0,1,...,T-1\}$:
			\begin{enumerate}
				\item Single source: $|\Tilde{V}|$ = $|\{\Tilde{v}\}| = 1$ and $V_r = \phi$.
				\item Single message induction: $m_0^{(\Tilde{v})} \neq \phi$ and  $m_{t > 0}^{(\Tilde{v})} = \phi$.
				\item Forward-and-forget: $L=1$, ${P_{f_1}}^{(v)}_t = \delta(t-t_R) \mathbbm{1}_{\{\mathcal{I}(f_1) \in \Theta\}}$, where $t_R$ is the time of arrival of the message $f_1$ into the feed.
				\item $P_{sp} = 0$ and $F^{(v)}_0 = m | \mathcal{I}(m) \in \Bar{\Theta}$.
			\end{enumerate}
		\end{definition}
		In other words, the message originated from node $\Tilde{v}$ at time $t=0$, performs a walk on the graph $G$. Considering this model, we state the following proposition:
		\begin{proposition}
			Let $\left(a^{(v_t)}_t\right)_{0\leq t\leq T-1}$ be a sequence of actions taken by the sequence of nodes $(v_t)_{0\leq t\leq T-1}$ where $a^{(v_t)}_t = v_{t+1}$ and $v_t \neq v_{t+1}$, $\forall t \in \{0,1,...,T-1\}$. Let the initial opinion of any regular node $v \in V_R$ about the class $\Tilde{\theta}$ associated with the source $\Tilde{v}$ be $\mu_{\Tilde{\theta},0}^{(v)}$ and let $\mu_{\Tilde{\theta},t}^{(v)}$ be the opinion at time $t$. For the \textit{simplified communication model}, maximizing the total opinion of the population at time $T$, i.e., $\sum\limits_{v \in V} \mu^{(v)}_{\Tilde{\theta},T}$ is equivalent to maximizing the sum, $\sum_{t=0}^{T-1} \mu_{\Tilde{\theta},t}^{(v_{t+1})} (1-\mu_{\Tilde{\theta},t}^{(v_{t+1})}) \left( \mu_{\Tilde{\theta},t}^{(v_{t+1})} + \alpha^{(v_{t+1})}_{\Tilde{\theta}} \beta^{(v_{t+1})}/\zeta^{(v_{t+1})}\right)^{-1}$.
		\end{proposition}
		\textit{Proof:}
		\begin{equation}
		\underset{\left(a^{(v_t)}_t\right)_{0\leq t\leq T-1}}{\text{maximize}} \sum\limits_{v \in V} \mu^{(v)}_{\Tilde{\theta},T}
		\equiv \underset{\left(a^{(v_t)}_t\right)_{0\leq t\leq T-1}}{\text{maximize}} \sum\limits_{v \in V} \mu^{(v)}_{\Tilde{\theta},T} - \mu^{(v)}_{\Tilde{\theta},0}.
		\end{equation}
		The objective function can be rewritten as follows:
		\begin{flalign}
		\sum\limits_{v \in V}& \mu^{(v)}_{\Tilde{\theta},T} - \mu^{(v)}_{\Tilde{\theta},0} = \sum\limits_{v \in V} \sum\limits_{t=0}^{T-1}\left( \mu^{(v)}_{\Tilde{\theta},t+1} - \mu^{(v)}_{\Tilde{\theta},t} \right) \notag & \\
		= \sum\limits_{t=0}^{T-1} & \left( \mu^{\left(a_t^{(v_t)}\right)}_{\Tilde{\theta},t+1} - \mu^{\left(a_t^{(v_t)}\right)}_{\Tilde{\theta},t} \right) + \sum\limits_{v^\prime \in V} \hspace{-0.2cm} \sum\limits_{\substack{t=0 \\ t:v^\prime \neq a_t^{(v_t)}}}^{T-1} \hspace{-0.2cm} \left( \mu^{(v^\prime)}_{\Tilde{\theta},t+1} - \mu^{(v^\prime)}_{\Tilde{\theta},t} \right).  \hspace{-0.5cm} &
		\label{eq:prop1_eq2}
		\end{flalign}
		
		It can be noticed that the second summand vanishes, since the summation is over only those nodes and time steps where there is no change in the opinion. Considering the term under the first summation, and substituting for $\mu_{\Tilde{\theta},t}^{(v_{t+1})}$ and $\mu_{\Tilde{\theta},t+1}^{(v_{t+1})}$ from Definition \ref{def:opin} and Eq.~\eqref{eq:dirichletPredict}, and omitting the superscript without loss of generality, $\forall t \in \{0,1,...,T-1\}$, we get:
		\begin{flalign}
		\mu_{\Tilde{\theta},t+1} &- \mu_{\Tilde{\theta},t} = \frac{\beta \alpha_{\Tilde{\theta}, t} + \zeta}{\beta \rho_t + \zeta} - \frac{\alpha_{\Tilde{\theta}, t}}{\rho_t} = \frac{\zeta  ({\rho_t} - \alpha_{\Tilde{\theta}, t})}{\left(\beta {\rho_t} + \zeta \right) {\rho_t} } \notag &
		\\
		&= \mu_{\Tilde{\theta},t} (1-\mu_{\Tilde{\theta},t}) \left(\mu_{\Tilde{\theta},t} + \alpha_{\Tilde{\theta}, t} \beta/\zeta \right)^{-1} .
		\end{flalign}
		Substituting the aforementioned equation in Eq.~\eqref{eq:prop1_eq2}, we conclude that maximizing $\sum\limits_{v \in V} \mu^{(v)}_{\Tilde{\theta},T}$ is equivalent to maximizing the following over $\left((v_{t+1})\right)_{0\leq t\leq T-1}$:
		\begin{equation}
		\sum\limits_{t=0}^{T-1} \mu_{\Tilde{\theta},t}^{\hspace{0.1mm}^{(v_{t+1})}} (1-{\mu_{\Tilde{\theta},t})^{\hspace{0.1mm}^{(v_{t+1})}}} \left( \mu_{\Tilde{\theta},t}^{\hspace{0.1mm}^{(v_{t+1})}} + \alpha^{\hspace{0.1mm}^{(v_{t+1})}}_{\Tilde{\theta}} \beta^{\hspace{0.1mm}^{(v_{t+1})}}/\zeta^{\hspace{0.1mm}^{(v_{t+1})}}\right)^{-1}.
		\qedsymbol 
		\label{eq:prop1last}
		\end{equation}
		\section{Proof of Proposition \ref{prop:toymodel}}
		\label{app:toymodelproof}
		Given the condition on the ratio of individual rewards $\frac{r_d}{r_c}$, first we shall prove the inequality $R_{cd} > R_{cc} > R_{dd}$, based on which the optimal mixed strategy is obtained.
		\subsection{Proving $R_{cd} > R_{cc}$:}
		Using the inequality $\frac{1}{ 1 + 2 \eta^{(c)} } < \frac{r_d}{r_c}$ we proceed as follows:
		\begin{equation}
		\left(\frac{\beta^{(c)} \rho^{(c)}}{ \beta^{(c)} \rho^{(c)} + 2 \zeta^{(c)}  } \right) r_c < r_d.
		\end{equation}
		Adding $r_c$ to both sides, we get:
		\begin{equation}
		\left( \frac{\beta^{(c)} \rho^{(c)}}{ \beta^{(c)} \rho^{(c)} + 2 \zeta^{(c)} }  + 1 \right) r_c  <  r_d + r_c.
		\end{equation}
		Substituting for $r_c$:
		\begin{equation}
		\left( \frac{\alpha_{\theta_2}^{(c)} \zeta^{(c)} }{\beta^{(c)} \rho^{(c)} + \zeta^{(c)}} \right) \left( \frac{2 \left(\beta^{(c)} \rho^{(c)} + \zeta^{(c)}\right)}{\beta^{(c)} \rho^{(c)} + 2 \zeta^{(c)}} \right) < r_d + r_c.
		\end{equation}
		Simplifying and using the fact that $R_{cd} = r_c + r_d$, we obtain:
		\begin{equation}
		\left( \frac{2 \alpha_{\theta_2}^{(c)} \zeta^{(c)} }{\beta^{(c)} \rho^{(c)} + 2 \zeta^{(c)}} \right) = R_{cc} < R_{cd}.
		\end{equation}	
		
		\subsection{Proving $R_{cc} > R_{dd}$:}
		From the given inequality $\frac{r_d}{r_c} < \frac{1+\eta^{(c)}}{1+2 \eta^{(c)}} \frac{1+2 \eta^{(d)}}{1+\eta^{(d)}}$ we get:
		\begin{equation}
		\begin{split}
		\frac{\alpha_{\theta_2}^{(d)} \zeta^{(d)} (\beta^{(c)} \rho^{(c)} + \zeta^{(c)}) \rho^{(c)} }{(\beta^{(d)} \rho^{(d)} + \zeta^{(d)}) \rho^{(d)} \alpha^{(c)} \zeta^{(c)}} & \\
		< \left(\frac{\beta^{(c)} \rho^{(c)} + \zeta^{(c)}}{\beta^{(c)} \rho^{(c)} + 2 \zeta^{(c)}} \right) & \left(\frac{\beta^{(d)} \rho^{(d)} + 2 \zeta^{(d)}}{\beta^{(d)} \rho^{(d)} + \zeta^{(d)}} \right).
		\end{split}
		\end{equation}
		Rearranging and simplifying, we get:
		\begin{equation}
		\frac{2 \alpha_{\theta_2}^{(d)} \zeta^{(d)}}{\beta^{(c)} \rho^{(d)} + 2 \zeta^{(d)}} < \frac{2 \alpha_{\theta_2}^{(c)} \zeta^{(c)}}{\beta^{(c)} \rho^{(c)} + 2 \zeta^{(c)}},
		\end{equation}
		which is equivalent to:
		\begin{equation}
		R_{dd} < R_{cc}.
		\end{equation}
		
		Note that since $R_{cc} > R_{dd}$, taking best selfish actions (by nodes $x$ and $y$) yields the joint reward $R_{cc}$.  However, since $R_{cd} > R_{cc}$ there is scope for improvement. Next, we determine the optimal mixed strategy which yields a joint reward better than $R_{cc}$ (but not equal $R_{cd}$).
		
		\subsection{Obtaining the mixed strategy}
		Let ${\pi} = (p, \Bar{p})$ be the mixed strategy chosen by both nodes $x$ and $y$, where $p= P(a^{(x)} = c) = P(a^{(y)} = c)$. The expected reward obtained using strategy ${\pi}$ is given by:
		\begin{equation}
		\mathbbm{E}_{\pi}[r] = R_{cc} p^2 + 2 R_{cd} p \Bar{p} +  R_{dd} \Bar{p}^2.
		\end{equation} 
		Setting $\frac{d}{dp}(\mathbbm{E}_{\pi}[r]) = 0$, and solving for $p$ we get:
		\begin{equation}
		p^* = \frac{1}{1+\frac{R_{cd}-R_{cc}}{R_{cd} - R_{dd}}}.
		\end{equation}	 
		Thus, the maximum expected normalized reward is $E_{\pi^*}[r] = R_{cc}+\frac{(R_{cd}-R_{cc})^2}{2 R_{cd} - R_{cc}-R_{dd}} > R_{cc}$. This implies that the expected joint reward obtained by using the mixed strategy $\bm{\pi}^*$ is greater than the maximum joint reward obtained by selfish actions, which is $R_{cc}$. \qedsymbol
		
		\section{Deriving $\mathbbm{E}\left\{\Delta X^{(v)}_{\theta, \tau} \mid \bm{\alpha}_0, \mathbf{a}_0\right\}$ (Eq.~\eqref{eq:delblf})}
		\label{app:delblfderivn}
		\noindent Let $Z^{(u,v)}_{\theta, \tau}$ be a Bernoulli random variable representing the event that node $u \in V$ pushes a message of class $\theta \in \Theta$ to node $v \in V_R$ at time $\tau$ and the corresponding probability be denoted by $P^{(\theta)}_{\tau}\{u \rightarrow v \mid \bm{\alpha}_0, \mathbf{a}_0\}$. Let ${\Delta X_{\theta, \tau}^{(v)}}$ be the random variable that denotes the change in belief parameter $\alpha^{(v)}_{\theta, \tau}$. The expected change in the belief parameter is given by:
		\begin{equation}
		\begin{split}
		\mathbbm{E}\left[\Delta X^{(v)}_{\theta, \tau} \mid \bm{\alpha}_0, \mathbf{a}_0\right] = \zeta^{(v)}  \sum\limits_{u \in \mathcal{N}(v)} \mathbbm{E}(Z^{(u,v)}_{\theta, \tau} \mid \bm{\alpha}_0, \mathbf{a}_0) \\
		= \zeta^{(v)} \sum\limits_{u \in \mathcal{N}(v)} P^{\theta}_{\tau}\{u \rightarrow v \mid \bm{\alpha}_0, \mathbf{a}_0\},
		\end{split}
		\end{equation}
		where
		\begin{equation}
		\begin{split}
		P^{\theta}_\tau\{u \rightarrow v \mid \bm{\alpha}_0, \mathbf{a}_0\} = \hspace{4.8cm}\\
		\hspace{-6mm} \int_{X_\tau}^{} \hspace{-2mm} f\left( u \text{ Tx } m_u, A^{(v)}_\tau\hspace{-1mm} = v, \mathcal{I}(m_u) = \theta, X_\tau \mid \bm{\alpha}_0, \mathbf{a}_0 \right) dX_\tau.
		\end{split}	 
		\label{eq:Puvmarginaliz}
		\end{equation}
		Since $f(X_\tau \mid \bm{\alpha}_0, \mathbf{a}_0) = \delta\left( X_\tau -\hat{\bm{\alpha}}_\tau \right)$, and the integrand (probability density) in Eq.~\eqref{eq:Puvmarginaliz} is independent of $X_0$ and $A_0$ given $X_\tau$, we get:
		\begin{flalign}
		P^{\theta}_\tau & \{u \rightarrow v \mid \bm{\alpha}_0, \mathbf{a}_0\} \notag \\
		&= P\left( u \text{ Tx } m_u, A^{(v)}_\tau = v, \mathcal{I}(m_u) = \theta \mid \hat{\bm{\alpha}}_\tau \right) \notag \\
		&= P(\left( u \text{ Tx } m_u \mid \mathcal{I}(m_u) = \theta,  \hat{\bm{\alpha}}_\tau \right) \notag \\
		& P\left( A^{(v)}_\tau = v \mid \mathcal{I}(m_u) = \theta, \hat{\bm{\alpha}}_\tau \right) P(\mathcal{I}(m_u) = \theta \mid \hat{\bm{\alpha}}_\tau) \notag \\
		&= \mu^{(u)}_{\theta, \tau} \pi^{(u)}_{v, \tau} (1-P_{sp}) \delta(\theta - \hat{\bm{\omega}}^{(v)}_\tau).
		\end{flalign}
		
		\bibliographystyle{IEEEtran}
		\bibliography{opinmax}
		
	\end{document}